\documentstyle[12pt,epsf,epsfig,cite]{article}
 1
 1
 1

\begin{document}
\begin{titlepage}
\def\baselinestretch{1.2}
\topmargin     -0.25in

\vspace*{\fill}
\begin{center}
{\large 
{\bf  Effects of Non-Standard Tri-linear Couplings in Photon-Photon Collisions: 
I. \mbox{$\gamma \gamma \rightarrow W^+ W^- \;$} }}
\vspace*{0.5cm}

\begin{tabular}[t]{c}
 
{\bf M.~Baillargeon$^{1}$,  G.~B\'elanger$^{2}$  and F.~Boudjema$^{2}$ }
 \\
\\
\\
{\it 1.  Grupo Te\'orico de Altas Energias, Instituto Superior T\'ecnico}\\
{\it Edif\'{\i}cio Ci\^encia (F\'{\i}sica)
P-1096 Lisboa Codex, Portugal}\\
{\it 2. Laboratoire de Physique Th\'eorique} 
EN{\large S}{\Large L}{\large A}PP
\footnote{URA 14-36 du CNRS, associ\'ee \`a l'E.N.S de Lyon et \`a
l'Universit\'e de Savoie.}\\
{\it Chemin de Bellevue, B.P. 110, F-74941 Annecy-le-Vieux, Cedex, France.}
\end{tabular}
\end{center}

\centerline{ {\bf Abstract} }
\baselineskip=14pt
\noindent
 {\small The effect of anomalous couplings in $\gamma \gamma \rightarrow W^+ W^-$ 
is studied for different energies of the 
$\gamma \gamma$ mode of the next linear collider. The analysis  based on  the maximum likelihood method exploits 
the variables in the four-fermion semi-leptonic final state. Polarised 
differential cross sections based on the complete set of diagrams for 
these channels with the inclusion of anomalous couplings are used and compared 
to an approximation based on $\gamma \gamma \rightarrow W^+ W^-$  with full spin correlations. To critically 
compare these results with those obtained in $e^+ e^-$ we perform an analysis 
based on the complete calculation of the four-fermion semi-leptonic final state.
The anomalous couplings that we consider are derived from the next-to-leading 
order operators in the non-linear realisation of symmetry breaking. }
\vspace*{\fill}
 
\vspace*{0.1cm}
\rightline{ENSLAPP-A-636/97}
\rightline{FISIST/3-97/CFIF}
\rightline{{\large  Jan. 1997}}
\end{titlepage}
\baselineskip=18pt


\newcommand{\be}{\begin{equation}}
\newcommand{\beq}{\begin{equation}}
\newcommand{\eeq}{\end{equation}}
\newcommand{\ee}{\end{equation}}

\newcommand{\beqn}{\begin{eqnarray}}
\newcommand{\eeqn}{\end{eqnarray}}
\newcommand{\bea}{\begin{eqnarray}}
\newcommand{\ena}{\end{eqnarray}} 
\newcommand{\ra}{\rightarrow}
 
\newcommand{\su}{$ SU(2) \times U(1)\,$}
\newcommand{\gag}{$\gamma \gamma$ }
\newcommand{\gam}{\gamma \gamma }

\newcommand{\np}{Nucl.\,Phys.\,}
\newcommand{\pl}{Phys.\,Lett.\,}
\newcommand{\pr}{Phys.\,Rev.\,}
\newcommand{\prl}{Phys.\,Rev.\,Lett.\,}
\newcommand{\prep}{Phys.\,Rep.\,}
\newcommand{\zp}{Z.\,Phys.\,}
\newcommand{\sovjnp}{{\em Sov.\ J.\ Nucl.\ Phys.\ }}
\newcommand{\nuclinst}{{\em Nucl.\ Instrum.\ Meth.\ }}
\newcommand{\annp}{{\em Ann.\ Phys.\ }}
\newcommand{\intjmp}{{\em Int.\ J.\ of Mod.\  Phys.\ }}
 
\newcommand{\eps}{\epsilon}
\newcommand{\mw}{M_{W}}
\newcommand{\mww}{M_{W}^{2}}
\newcommand{\mwmw}{M_{W}^{2}}
\newcommand{\mhmh}{M_{H}^2}
\newcommand{\mz}{M_{Z}}
\newcommand{\mzz}{M_{Z}^{2}}

\newcommand{\lra}{\leftrightarrow}
\newcommand{\tr}{{\rm Tr}}
\def\ls1{{\not l}_1} 
\newcommand{\cms}{centre-of-mass\hspace*{.1cm}}
 
\newcommand{\dkg}{\Delta \kappa_{\gamma}}
\newcommand{\dkz}{\Delta \kappa_{Z}}
\newcommand{\dz}{\delta_{Z}}
\newcommand{\dgz}{\Delta g^{1}_{Z}}
\newcommand{\dgzt}{$\Delta g^{1}_{Z}\;$}
\newcommand{\la}{\lambda}
\newcommand{\lag}{\lambda_{\gamma}}
\newcommand{\laz}{\lambda_{Z}}
\newcommand{\lnl}{L_{9L}}
\newcommand{\lnr}{L_{9R}}
\newcommand{\lt}{L_{10}}
\newcommand{\lu}{L_{1}}
\newcommand{\ld}{L_{2}}
\newcommand{\cw}{\cos\theta_W}
\newcommand{\sw}{\sin\theta_W}
\newcommand{\tw}{\tan\theta_W}

\newcommand{\epm}{$e^{+} e^{-}\;$}
\newcommand{\epemt}{$e^{+} e^{-}\;$}
\newcommand{\epem}{e^{+} e^{-}\;}
\newcommand{\ememt}{$e^{-} e^{-}\;$}
\newcommand{\emem}{e^{-} e^{-}\;}
\newcommand{\eeww}{e^{+} e^{-} \ra W^+ W^- \;}
\newcommand{\eewwt}{$e^{+} e^{-} \ra W^+ W^- \;$}
\newcommand{\epemww}{e^{+} e^{-} \ra W^+ W^- }
\newcommand{\epemwwt}{$e^{+} e^{-} \ra W^+ W^- \;$}
\newcommand{\eennhht}{$e^{+} e^{-} \ra \nu_e \bar \nu_e HH\;$}
\newcommand{\eennhh}{e^{+} e^{-} \ra \nu_e \bar \nu_e HH\;}
\newcommand{\ppwg}{p p \ra W \gamma}
\newcommand{\wwhh}{W^+ W^- \ra HH\;}
\newcommand{\wwhht}{$W^+ W^- \ra HH\;$}
\newcommand{\ppwz}{pp \ra W Z}
\newcommand{\ppwgt}{$p p \ra W \gamma \;$}
\newcommand{\ppwzt}{$pp \ra W Z \;$}
\newcommand{\gamgamt}{$\gamma \gamma \;$}
\newcommand{\gamgam}{\gamma \gamma \;}
\newcommand{\egamt}{$e \gamma \;$}
\newcommand{\egam}{e \gamma \;}
\newcommand{\gamgamwwt}{$\gamma \gamma \ra W^+ W^- \;$}
\newcommand{\gamgamwwht}{$\gamma \gamma \ra W^+ W^- H \;$}
\newcommand{\gamgamwwh}{\gamma \gamma \ra W^+ W^- H \;}
\newcommand{\gamgamwwhht}{$\gamma \gamma \ra W^+ W^- H H\;$}
\newcommand{\gamgamwwhh}{\gamma \gamma \ra W^+ W^- H H\;}
\newcommand{\ggww}{\gamma \gamma \ra W^+ W^-}
\newcommand{\ggwwt}{$\gamma \gamma \ra W^+ W^- \;$}
\newcommand{\ggwwht}{$\gamma \gamma \ra W^+ W^- H \;$}
\newcommand{\ggwwh}{\gamma \gamma \ra W^+ W^- H \;}
\newcommand{\ggwwhht}{$\gamma \gamma \ra W^+ W^- H H\;$}
\newcommand{\ggwwhh}{\gamma \gamma \ra W^+ W^- H H\;}
\newcommand{\ggwwz}{\gamma \gamma \ra W^+ W^- Z\;}
\newcommand{\ggwwzt}{$\gamma \gamma \ra W^+ W^- Z\;$}
\def\smx{{\cal{S}} {\cal{M}}\;}

\newcommand{\ptu}{p_{1\bot}}
\newcommand{\vecptu}{\vec{p}_{1\bot}}
\newcommand{\ptd}{p_{2\bot}}
\newcommand{\vecptd}{\vec{p}_{2\bot}}
\newcommand{\ie}{{\em i.e.}}
\newcommand{\cm}{{{\cal M}}}
\newcommand{\cl}{{{\cal L}}}
\newcommand{\cd}{{{\cal D}}}
\newcommand{\cv}{{{\cal V}}}
\def\slashc{c\kern -.400em {/}}
\def\slashL{L\kern -.450em {/}}
\def\slashcl{\cl\kern -.600em {/}}
\def\Ww{{\mbox{\boldmath $W$}}}  
\def\B{{\mbox{\boldmath $B$}}}         
\def\noi{\noindent}
\def\nn{\noindent}
\def\sm{${\cal{S}} {\cal{M}}\;$}
\def\nph{${\cal{N}} {\cal{P}}\;$}
\def\sb{$ {\cal{S}}  {\cal{B}}\;$}
\def\ssb{${\cal{S}} {\cal{S}}  {\cal{B}}\;$}
\def\ssbe{{\cal{S}} {\cal{S}}  {\cal{B}}}
\def\cviol{${\cal{C}}\;$}
\def\pviol{${\cal{P}}\;$}
\def\cpviol{${\cal{C}} {\cal{P}}\;$}

\newcommand{\lgg}{\lambda_1\lambda_2}
\newcommand{\lww}{\lambda_3\lambda_4}
\newcommand{\ppin}{ P^+_{12}}
\newcommand{\pmin}{ P^-_{12}}
\newcommand{\ppout}{ P^+_{34}}
\newcommand{\pmout}{ P^-_{34}}
\newcommand{\sinsq}{\sin^2\theta}
\newcommand{\cossq}{\cos^2\theta}
\newcommand{\yt}{y_\theta}
\newcommand{\hppll}{++;00}
\newcommand{\hpmll}{+-;00}
\newcommand{\hpplt}{++;\lambda_30}
\newcommand{\hpmlt}{+-;\lambda_30}
\newcommand{\hpptt}{++;\lambda_3\lambda_4}
\newcommand{\hpmtt}{+-;\lambda_3\lambda_4} 
\newcommand{\dk}{\Delta\kappa}
\newcommand{\klam}{\Delta\kappa \lambda_\gamma }
\newcommand{\kac}{\Delta\kappa^2 }
\newcommand{\lac}{\lambda_\gamma^2 }
\def\gamgamtzz{$\gamma \gamma \ra ZZ \;$}
\def\gamgamtww{$\gamma \gamma \ra W^+ W^-\;$}
\def\gamgamtwwe{\gamma \gamma \ra W^+ W^-}

\setcounter{section}{1}

\setcounter{subsection}{0}
\setcounter{equation}{0}
\def\thesubsection {\thesection.\arabic{subsection}}
\def\theequation{\thesection.\arabic{equation}}
\setcounter{equation}{0}
\def\thequation{\thesection.\arabic{equation}}
\setcounter{section}{0} 
\setcounter{subsection}{0}
\section{Introduction}
The next linear \epemt collider\cite{NLCgeneral} can be turned   into a \gag
collider\cite{PhotonCol} by  
converting the single pass electrons into very energetic photons through Compton backscattering 
of a laser light,
whereby the obtained photon can take as much as $80\%$ of the initial beam energy. 
The main  attractions of such a  mode of the next linear  collider 
rest on its ability to study in detail the properties of a Higgs
\cite{Gunionzz,Gunioncphiggsgg,Peterparity,Gounarishiggs} that can be 
produced as a resonance, since two photons can be 
in a $J_Z=0$ state whereas chirality very much suppresses this configuration with 
the  \epemt pair. Also, because cross sections for the production of 
weak vector bosons are much larger 
in the \gag mode than in \epemt\cite{Parisgg}, the  very large samples of $W$'s could allow for high precision 
measurements on the properties of these gauge bosons. 
Considering that the  physics of $W$'s could reveal much on the 
dynamics of the Goldstone bosons through the longitudinal component of  the $W$'s 
this mode of the linear collider may appear ideal for an investigation of the mechanism 
of symmetry breaking. \\
 However, it is also true that these important issues can be easily 
blurred by  backgrounds that are often quite large in the photon mode. 
For instance, in the case of  the Higgs, a resonant signal is viable 
only for a light Higgs after judiciously tuning the parameters (energies and polarisations) 
of the \gag collider\cite{ggzzsm,Halzen,nousggHiggs}. The aim of the present paper is to critically 
analyse the extent to which 
the reaction   
\ggwwt can be useful in measuring the electromagnetic couplings of the $W$ and how these 
measurements compare to those one could perform in the ``natural" \epemt mode of 
the next collider. From the outset,  one would naively expect the \gag mode to fare much better 
than the \epemt mode when it comes to the anomalous $WW\gamma$ couplings, not only because 
the $WW$ statistics is much larger in the photon mode but also because the most 
promising reaction in \epemt, \eewwt, accesses not only $WW\gamma$ but also the $WWZ$ couplings. 
If these two types of couplings were not related to each other, as one has generally assumed, 
then it is 
quite difficult to disentangle  a $WW\gamma$ and a $WWZ$ coupling 
at the \epemt mode, whereas obviously only the former can be probed at the \gag mode. 
However, as we will
argue, the electroweak precision measurements at the $Z$ peak are a sign that there should be a hierarchy
of couplings 
whereby the symmetries one has observed at the present energies indicate that the $WW\gamma$ and the 
$WWZ$ should be related. If this is so,
though \ggwwt is unique in unambiguously measuring the $WW\gamma$ couplings, 
it is somehow probing the same parameter space as  \eewwt. In  such 
an eventuality one should then 
enquire about how to exploit the 
\gag mode, and whether the reaction \ggwwt can give more stringent constraints 
than in the \epemt mode. In addition it is useful to investigate 
whether one may gain by combining the results 
of the \gag analysis with those obtained in the \epemt mode. 
These two aspects will be addressed in the present paper. \\ 

Although there have been numerous studies that
dealt with the subject of the tri-linear anomalous couplings in \ggwwt\cite{nousggvvref}, they have all been conducted at
the level of the \ggwwt cross section. As we will show, even if one assumes reconstruction of the 
helicities of the $W$, restricting the analysis of the \ggwwt 
at the level of the  cross section, where one only accesses the diagonal
elements of the $WW$ density matrix, it is not possible to maximally enhance the effect of the
anomalous couplings. Indeed, as one expects in an investigation of the Goldtsone sector, the new 
physics parameterised in this context by an anomalous magnetic moment of the $W$, $\Delta
\kappa_\gamma$, affects principally the production of two longitudinal $W$ bosons. In the 
\gag mode this affects predominantly the $J_Z=0$ amplitude by providing an {\em enhanced
coupling} of the order $\gamma=s_{\gamma \gamma}/{M_W^2}$ ($\sqrt{s_{\gamma \gamma}}$ is the $\gamma
\gamma$ centre of mass energy). Unfortunately the same standard amplitude 
(the $J_Z=0$ with two longitudinal W's) 
has the factor $1/\gamma$ and therefore the interference is not effective 
in the sense that the genuine enhanced coupling $\gamma$ brought about by the new physics is washed out. This is in
contrast with what happens in the \epemt mode where the interference is fully effective. Nonetheless, the enhanced coupling 
 could still be  exploited in the \gag mode if one is able to reconstruct the 
{\em non diagonal} elements of the $WW$ density
matrix. This can be done by analysing the distributions involving kinematical variables 
of the decay products of the $W$. 
In any case,  in a realistic set-up the $W$'s are only reconstructed from their decay
products and since we would need to impose cuts on the fermions, one absolutely requires to have at
hand the distributions of the fermions emerging from the $W$'s. One thus needs the fully polarised 
$WW$ density matrix elements which one combines  with the polarised decay functions in order to 
keep the full spin correlations and arrive at a more precise description of \ggwwt in terms of 
$\gamma \gamma \ra W^+ W^- \ra 4f$. Having access to all 
the kinematics of the fermionic final states, one can exploit the powerful 
technique of the maximum likelihood method, ML, to search for an anomalous behaviour that can affect 
any of the distributions of the 4-fermion final state and in our case unravel the 
contribution of the non-diagonal elements of the $WW$ density matrix which are most sensitive 
to the anomalous couplings. The exploitation of the density matrix elements  in 
\eewwt has been found to be a powerful tool not only at LEP2 energies
\cite{Fernandeeww,Sekulin,wwanolep2}
but also at the next collider\cite{Barklowfits,Coutureee4fano,Moijapan,Gounariserato}, however a thorough investigation
in \gag is missing. \\
\noindent In a previous paper\cite{enslapp635}, dedicated to four fermion final states in \gag
within the \sm, we have shown that these signatures could be very well approximated 
by taking into account only the $WW$ resonant diagrams provided these were computed through 
the density matrix formalism and a smearing factor taking into account the finite width 
of the $W$ is applied. In this paper we will not  only consider the fully correlated 
$WW$ cross sections leading to a semi-leptonic final state  taking into account the anomalous
couplings but we will also consider the full set of the four-fermion final states including those 
anomalous couplings, thus avoiding any possible bias.
Having extracted the limits on the anomalous couplings in \gag we will contrast them with 
those one obtains in \epemt. For the latter we conduct our own analysis based on the same 
set of parameters as in the \gag mode and most importantly taking into account the full 
set of four fermion diagrams. The maximum likelihood method is used throughout. \\
The anomalous couplings that we study in this paper are derived from 
 a chiral Lagrangian formulation which does not require the  
Higgs\cite{chiralbib}. 
To critically compare the performance of the \epemt and the \gag mode, we will first consider 
the case where a full SU(2) global symmetry is implemented as well as a situation where one 
allows a breaking of this symmetry. We will see that the advantages of the \gag mode depend 
crucially on the model considered. 
\\
Our paper is organised as follows. After a brief motivation of the chiral Lagrangian and a
presentation of the operators that we want to probe, we give in section~3 a full
description of the helicity amplitudes and of the density matrix for \ggwwt including the 
anomalous couplings. We then proceed to extract limits on the couplings by exploiting the maximum
likelihood method both for the ``resonant" diagrams as well as for the complete set of Feynman
diagrams for the 
4-fermion final state, for various combinations of the photon helicities. 
We then discuss the limits one obtains in \epemt and compare them with those one obtains in \gag 
for different centre-of-mass energies. The last section contains our conclusions.

\section{Anomalous couplings and the chiral Lagrangian}
\setcounter{secnumdepth}{2} 
\setcounter{equation}{0}
\def\thesubsection{\thesection.\arabic{subsection}} 
\def\theequation{\thesection.\arabic{equation}} 
If by the time the next linear collider is built and if after 
a short run there has been no sign of a Higgs, one would have learnt 
that the supersymmetric extension 
of the \sm may not be realised, at least in its simplest 
form, and that the weak vector bosons may become strongly 
interacting. In this scenario, in order to probe the mechanism 
of symmetry breaking it will be of utmost importance to 
scrutinise the dynamics of the 
weak vector bosons since their longitudinal modes stem from the 
Goldstones bosons which are the remnants of the symmetry breaking sector. 
Well before the opening up of new thresholds one  expects the dynamics of the symmetry 
breaking sector 
to affect the self-couplings of the gauge bosons. The natural framework to describe, 
in a most general way, 
the physics that make do with a Higgs and that parameterises these self-couplings 
relies on an effective Lagrangian adapted from 
pion physics where the symmetry breaking is non-linearly realised
\cite{chiralbib}. 
This effective Lagrangian incorporates all the symmetries which have been 
verified so far, especially the gauge symmetry. 
We will assume that the gauge symmetry is \su. Moreover present 
precision measurements indicate that the $\rho$ parameter is such that 
$\rho \sim 1$~\cite{Li-Altarelli96}. This suggests that the electroweak 
interaction has 
a custodial global SU(2) symmetry
(after switching off the gauge couplings), whose slight breaking 
seems to be entirely due to the bottom-top mass splitting. This additional symmetry may 
be imposed on the effective Lagrangian. 
We will also assume that \cpviol is an exact symmetry. 
These  
ingredients should be incorporated when constructing the effective 
Lagrangian. The construction and approach has, lately, become widespread
in discussing weak bosons anomalous couplings in 
the absence of a Higgs. \\
The effective Lagrangian is organised 
as a set of operators whose leading order operators (in an energy expansion) 
reproduce the ``Higgsless" standard 
model. Introducing our notations, as concerns the purely 
bosonic sector, the SU(2) kinetic term that gives the standard 
tree-level gauge self-couplings is
\beqn
\cl_{{\rm Gauge}}=- \frac{1}{2} \left[ 
\tr(\Ww_{\mu \nu} \Ww^{\mu \nu}) + \tr(\B_{\mu \nu} \B^{\mu \nu}) \right]
\eeqn

where the $SU(2)$ gauge fields  are $\Ww_\mu=W^{i}_{\mu}\tau^i$, 
while the hypercharge 
field is denoted by $\B_\mu=\tau_3 B_\mu$. The normalisation for the Pauli 
matrices is $\tr(\tau^i \tau^j)=2\delta^{i j}$. 
We define the field strength as,
$\Ww_{\mu \nu}$ 
\beqn
\Ww_{\mu \nu}&=&\frac{1}{2} \left( 
\partial_\mu \Ww_{\nu}- \partial_\nu \Ww_{\mu} +\frac{i}{2}
g[\Ww_\mu, \Ww_\nu] 
\right) \nonumber \\
&=&\frac{\tau^i}{2} \left(\partial_\mu W^{i}_\nu-\partial_\nu W^{i}_\mu
-g \epsilon^{ijk}W_{\mu}^{j}W_{\nu}^{k} \right) 
\eeqn

The  Goldstone bosons, $\omega^i$, within the built-in SU(2) symmetry are 
assembled 
in a matrix $\Sigma$
\beqn
\Sigma=exp(\frac{i \omega^i \tau^i}{v}) \;\; ; v=246~GeV\;\;
\;\mbox{ and} \;\;
{{\cal D}}_{\mu} \Sigma=\partial_\mu \Sigma + \frac{i}{2} 
\left( g \Ww_{\mu} \Sigma
- g'B_\mu \Sigma \tau_3 \right)
\eeqn

This leads to the gauge invariant mass term for the $W$ and $Z$

\beqn
\cl_M=\frac{v^2}{4} \tr(\cd^\mu \Sigma^\dagger \cd_\mu \Sigma)
\equiv - \frac{v^2}{4}  \tr\left( \cv_\mu \cv^\mu \right) \;\;
\;\;\;\;
\cv_\mu=\left( {{\cal D}}_{\mu} \Sigma \right) \Sigma^{\dagger}
\;\; ; \;\;M_W=\frac{g v}{2}
\eeqn

The above operators are the leading order operators in conformity with 
the \su gauge symmetry and which incorporate the custodial SU(2) symmetry. 
They thus represent the minimal Higgsless electroweak model. At the same order 
we may include a breaking of the global symmetry through 

\beqn
\cl_{\Delta \rho}= \Delta \rho \frac{v^2}{8} 
\left( \tr ({{\cal V}}_\mu X ) \right)^2 
\;\;\;;\;\; \;X=\Sigma \tau^3 \Sigma^\dagger
\eeqn

\noi Global fits from the present data give\cite{Li-Altarelli96}, after having subtracted the 
\sm contributions\footnote{
These were evaluated with $m_t=175GeV$ and to keep within the spirit of 
Higgless model, $M_H=1$TeV, thus defining $\Delta \rho_{New}$.} 
\beq
-.1\;<\;10^{3}\; \Delta \rho_{New}\;<\;2.5  
\eeq

Different scenarios of New Physics connected with symmetry breaking 
are described by the Next-to-Leading-Order (NLO) operators. 
Maintaining the custodial symmetry only a few  operators are 
possible

\beqn
\label{NLOcustodial}
\cl_{NLO}=
&+& g g' \frac{L_{10}}{16\pi^2} 
\tr (\Sigma \B^{\mu\nu}\Sigma^\dagger \Ww_{\mu\nu} ) \nonumber \\
&-&i g' \frac{L_{9R}}{16 \pi^2} 
\tr ( {\bf B}^{\mu \nu}\cd_{\mu}
\Sigma^{\dagger} \cd_{\nu} \Sigma )
-i g \frac{L_{9L}}{16 \pi^2} \tr ( \Ww^{\mu \nu}\cd_{\mu}
\Sigma \cd_{\nu} \Sigma^{\dagger} )  \nonumber \\ 
&+&\frac{L_1}{16 \pi^2} \left( \tr (D^\mu \Sigma^\dagger D_\mu \Sigma) 
\right)^2 
+\frac{L_2}{16 \pi^2} \left( \tr (D^\mu \Sigma^\dagger D_\nu \Sigma)
\right)^2 
\eeqn

Some important remarks are in order. Although these are New Physics 
operators that should exhibit the corresponding new scale 
$\Lambda$, such a scale does not appear in our definitions. 
Implicitly, one has in Eq.~\ref{NLOcustodial} $\Lambda=4\pi v \simeq 3.1TeV$.  
The first operator $L_{10}$ contributes 
directly at tree-level to the two-point functions. The latter 
are extremely 
well measured at LEP1/SLC. Indeed $L_{10}$ is directly related to the new physics 
contribution to the 
Peskin-Takeuchi parameter\cite{Peskin}, $S_{{\rm new}}$: 
$L_{10} =-\pi S_{{\rm new}}$. The present inferred value of $L_{10}$ 
is  $-1.2 \leq L_{10} \leq 0.1$\cite{Li-Altarelli96}, after allowing for the \sm contribution. 
We will hardly improve on this limit 
in future experiments through double pair production, like \eewwt and \ggwwt. 
Therefore in the rest of the analysis we will set $L_{10}=0$ 
\footnote{It is possible to associate the vanishing of $L_{10}$ to an 
extra global symmetry\cite{InamiL10} in the same way that $\rho=1$ can be 
a reflection of the custodial symmetry. Extended BESS \cite{ExtendedBess} 
implements such a symmetry. We could have very easily included $L_{10}$ in our 
analysis, however our results show that bounds of order 
 ${{\cal O}}(.1)$ 
on $L_{9}$ will only be possible with a 2TeV machine. At this energy we 
may entertain the idea of constraining $L_{10}$ further than the existing 
limits from LEP1, however we will not pursue this here.}
and enquire whether 
one can set constraints of this order on the remaining parameters. If so, this will 
put extremely powerful constraints on the building of possible models of symmetry breaking.   
%
The two last operators, $L_1$ and $L_2$  
are the only ones that remain upon switching off the gauge couplings.  
They give rise to genuine quartic couplings that involve 
four predominantly longitudinal states. Therefore, phenomenologically, 
these would be the most likely to give large effects. Unfortunately 
they do not contribute to \ggwwt nor to \eewwt. Within the constraint of SU(2) 
global symmetry one is left with only the two operators 
$L_{9L,9R}$.

\noi
Both $L_{9L}$ and $L_{9R}$ induce \cviol and \pviol
 conserving $\gamma WW$ and
$ZWW$ couplings. 
In fact, for $WW\gamma$ both operators contribute equally to the same Lorentz structure 
and thus there is no way that \ggwwt could differentiate between these two operators. \\
By allowing for custodial symmetry breaking terms more operators are possible. 
A ``naturality argument" would suggest that the coefficients of these 
operators should be suppressed by a factor $\Delta \rho$ compared 
to those of $L_9$, following the observed hierarchy in the leading 
order operators. 
One of these operators(\cite{Feruglio},\cite{AppelquistWu}) stands out, because it leads to \cviol and \pviol 
violation and only affects the $WWZ$ vertex, and hence only contributes to 
\eewwt: 
\beqn
\slashcl_\slashc = g \frac{L_c }{16 \pi^2} 
\left( \tr ( \widetilde{\Ww}^{\mu \nu} \cv_\mu ) \right) 
\left( \tr (X \cv_\nu ) \right) \;\;\; ;\;\;
\widetilde{\Ww}^{\mu \nu}=\frac{1}{2} 
\eps^{\mu \nu \alpha \beta} \;\Ww_{\alpha \beta}
\eeqn
%

For completeness we will also consider another operator which breaks this global 
symmetry without leading to any \cviol and \pviol breaking: 

\beqn
\slashcl_1 =i g \frac{\slashL_1}{16 \pi^2} 
\left( \tr ( \Ww^{\mu \nu} X ) \right) 
\left( \tr (X [\cv_\mu,\cv_\nu] ) \right)
\eeqn

It has become customary to refer to the popular phenomenological 
parametrisation (the {\em HPZH} parameterization)\cite{HPZH} that 
gives the most general tri-linear coupling  that could contribute 
to \eewwt. We reproduce it here  to show how the above 
chiral Lagrangian operators  show up in \eewwt and  \ggwwt. 
The phenomenological 
parametrisation writes 
\beqn
{\cal L}_{WWV}&=& -ie \left\{ \left[ A_\mu \left( W^{-\mu \nu} W^{+}_{\nu} - 
W^{+\mu \nu} W^{-}_{\nu} \right) +  
\overbrace{ (1+\mbox{\boldmath $\Delta \kappa_\gamma$} )}^{\kappa_\gamma} 
F_{\mu \nu} W^{+\mu} W^{-\nu} \right] \right.
\nonumber \\
&&+ \left. cotg \theta_w \left[\overbrace{(1+ {\bf \Delta g_1^Z})}^{g_1^Z}
Z_\mu \left( W^{-\mu \nu} W^{+}_{\nu} - 
W^{+\mu \nu} W^{-}_{\nu} \right) +
\overbrace{(1+\mbox{\boldmath $\Delta \kappa_Z$} )}^{\kappa_Z} 
Z_{\mu \nu} W^{+\mu} W^{-\nu} \right] \right.
\nonumber \\
&&+ \left. \frac{1}{M_{W}^{2}} 
\left( \mbox{\boldmath $\lambda_\gamma$} \;F^{\nu \lambda}+
\mbox{\boldmath $\lambda_Z$} \;
cotg \theta_w Z^{\nu \lambda}
\right) W^{+}_{\lambda \mu} W^{-\mu}_{\;\;\;\;\;\nu} \right\} \\ \nonumber 
&-& e \frac{c_W}{s_W}  g_5^Z \left\{
\epsilon^{\mu \nu \rho \sigma} 
\left(W_\mu^+ (\partial_\rho W_\nu) - 
(\partial_\rho W_{\mu}^{+}) W_\nu \right)Z_\sigma  \right\}
\eeqn
 To map the operators 
we have introduced into this parameterisation one needs to specialise to 
the unitary gauge by setting the Goldstone ($\omega_i$) fields to zero 
($\Sigma \ra 1$). We find  
\beqn \label{constraints}
\Delta\kappa_\gamma&=&
\frac{e^2}{s_w^2} \frac{1}{32 \pi^2} \left( L_{9L}+L_{9R} + 4\slashL_1\right)
\equiv \frac{e^2}{s_w^2} \frac{1}{32 \pi^2} L_\gamma
\nonumber \\
\Delta\kappa_Z&=&
\frac{e^2}{s_w^2} \frac{1}{32 \pi^2} \left( L_{9L} 
-\frac{s_w^2}{c_w^2} L_{9R} + 4\slashL_1 \right) 
\nonumber \\
\Delta g_1^Z&=&
\frac{e^2}{s_w^2} \frac{1}{32 \pi^2} \left(\frac{ L_{9L}}{c_w^2} \right)
\nonumber \\
g_5^Z&=&\frac{e^2}{32 \pi^2 \sin^2\theta_w} 
\left(-\frac{L_c}{\cos^2\theta_w}\right) \nonumber \\
\lambda_\gamma&=&\lambda_Z=0 
\eeqn

It is important to note that within the Higgless implementation of the ``anomalous" couplings 
the $\lambda_{\gamma, Z}$ are not induced at the next-to-leading order. 
They represent weak bosons 
that are essentially transverse and therefore do not efficiently probe the symmetry breaking 
sector as evidenced by the fact that they do not involve the Goldstone bosons. 
It is worth remarking that, in effect, within the chiral Lagrangian approach 
there is essentially only one
 effective coupling parameterised by $\Delta \kappa_\gamma$ 
that one may reach in \gag.
Thus \ggwwt probes the collective combination $L_\gamma\equiv L_{9L}+L_{9R} + 4\slashL_1$. 
However the four independent operators contribute differently to the various 
Lorentz structure of $WWZ$ and thus to \epemwwt.

For completeness here are the quartic couplings that accompany the tri-linear 
parts as derived from the chiral effective Lagrangian

\beqn
\cl^{SM}_{WWV_{1}V_{2}} &=& -e^2 \left\{ \left(
A_\mu A^\mu W^{+}_{\nu} W^{- \nu} - A^\mu A^\nu W^{+}_{\mu} 
W^{-}_{\nu} \right) \right. \nonumber \\
&+& 2 \frac{c_w}{s_w} (1+\frac{l_{9l}}{c_w^2}) \left(
A_\mu Z^\mu W^{+}_{\nu}W^{-\nu} - \frac{1}{2} 
A^\mu Z^\nu ( W^{+}_{\mu}W^{-}_{\nu} + W^{+}_{\nu}W^{-}_{\mu} ) \right) 
\nonumber \\
&+&\frac{c_w^2}{s_w^2} (1+\frac{2 l_{9l}}{c_w^2}-\frac{l_-}{c_w^4}) \left(
Z_\mu Z^\mu W^{+}_{\nu}W^{-\nu} - Z^\mu Z^\nu W^{+}_{\mu}W^{-}_{\nu} \right) 
\nonumber \\
&+& \frac{1}{2 s_w^2} (1+2 l_{9l}-l_-) 
 \left(W^{+\mu} W^{-}_{\mu} W^{+\nu} W^{-}_{\nu} -
W^{+ \mu} W^{+}_{\mu} W^{-\nu}W^{-}_{\nu} \right) \nonumber \\
&-&\frac{l_+}{2s_w^2}\left( \left( 3 W^{+\mu} W^{-}_{\mu} W^{+\nu} W^{-}_{\nu}+
W^{+ \mu} W^{+}_{\mu} W^{-\nu}W^{-}_{\nu} \right) \right. \nonumber \\
&+&\frac{2}{c_w^2} \left. \left. \left(Z_\mu Z^\mu W^{+}_{\nu}W^{-\nu} + Z^\mu Z^\nu W^{+}_{\mu}W^{-}_{\nu}
 \right) + \frac{1}{c_w^4} Z_\mu Z^\mu  Z_\nu Z^\nu \right)\right. \nonumber \\
&-& \left. 2 i \frac{c_W}{s_W} g_5^Z \varepsilon^{\mu \nu \alpha \beta} 
A^{\mu} Z^{\nu} W^{+}_{\alpha} W^{-}_{\beta} \right\}\nonumber \\
 {\rm with} &&
(l_{9l},r_9,\tilde{l}_1)=\frac{e^2}{32 \pi^2 s_w^2} 
(L_{9L},L_{9R},\slashL_1)\;\;\; ;\;\;\;l_\pm=\frac{e^2}{32 \pi^2 s_w^2}
(L_1 \pm L_2)
\eeqn



\section{Characteristics of helicity amplitudes for \ggwwt\\
and comparison with \eewwt} 
\setcounter{secnumdepth}{2} 
\setcounter{equation}{0}
\def\thesubsection{\thesection.\arabic{subsection}} 
\def\theequation{\thesection.\arabic{equation}} 
\subsection{\ggwwt}
\begin{figure}
  \vspace*{-1.5cm}
  \begin{center}
    \mbox{\epsfxsize=16cm\epsfysize=9cm\epsffile{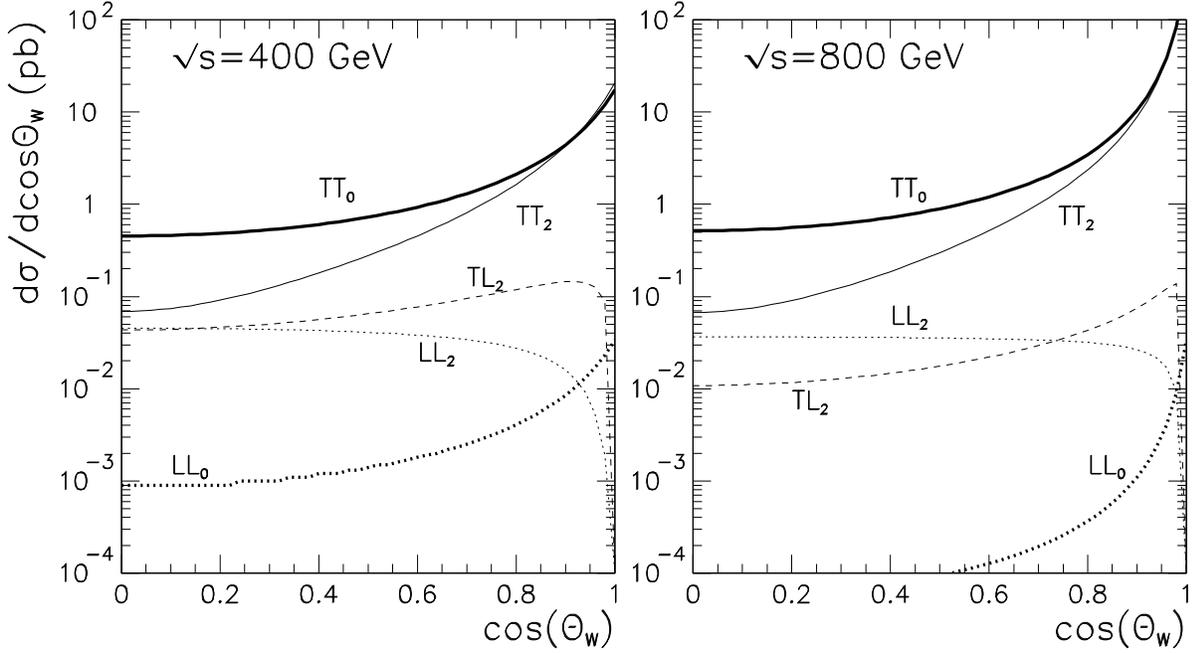}}
  \end{center}
  \vspace*{-.9cm}
\caption{\label{wwsmdis} {\em Angular distribution for different polarisation 
of the $W$'s in \ggwwt at $\sqrt{s}=400$ and $800$GeV.}}
  \vspace{1cm}
\end{figure}

\begin{figure}
  \vspace*{-1.5cm}
  \begin{center}
    \mbox{\epsfxsize=16cm\epsfysize=9cm\epsffile{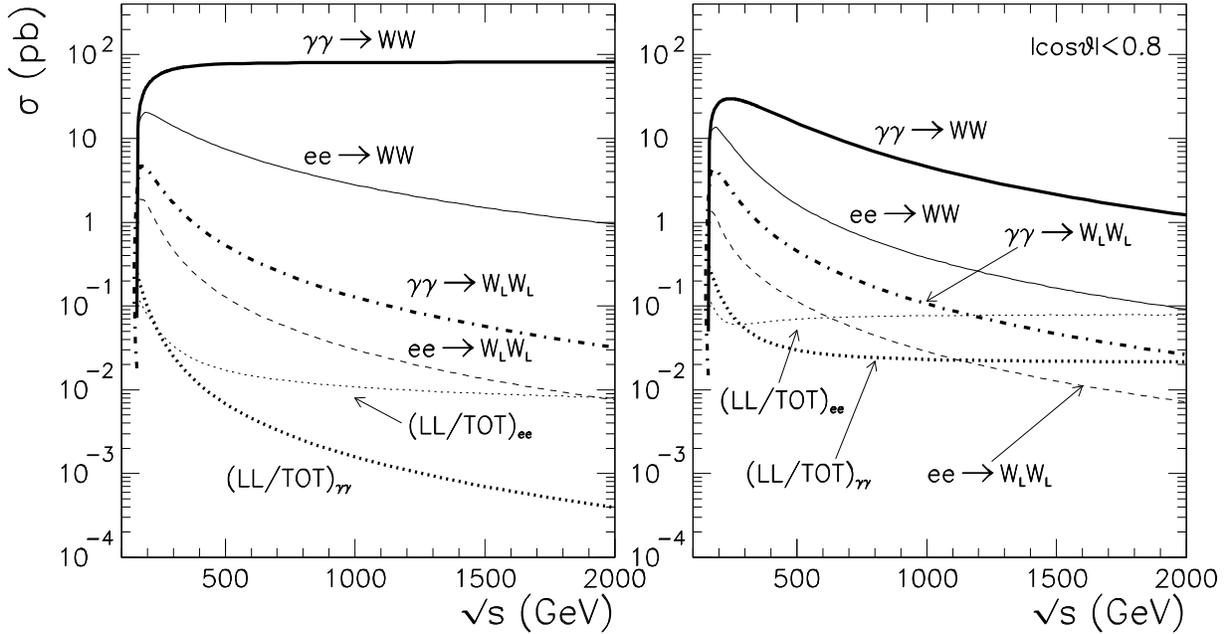}}
  \end{center}
  \vspace*{-.9cm}
\caption{\label{eeggww} 
{\em Energy dependence of the \ggwwt versus \epemwwt 
cross sections, for different polarisation. The ratio of longitudinal over transverse 
is shown in both cases. The effect of a cut on the scattering angle is also shown.}}
  \vspace{0cm}
\end{figure}

It is very instructive to stress some very simple but important properties of the \ggwwt 
differential and total cross section in the \sm, since this will greatly help 
in devising the best strategy to maximise the effects of the chiral Lagrangian 
operators.  
The characteristics of the \ggwwt cross section are 
most easily revealed in the expression of the helicity amplitudes. We have derived 
these (see Appendix~A) in a very compact form, both in the \sm and in the 
presence of the anomalous couplings. 
The characteristics of the helicity amplitudes are drastically different in the two cases. 
As concerns the \sm, 
the bulk of the cross section is due to forward $W$'s (see Fig.~\ref{wwsmdis}). 
More importantly,  
the cross section is dominated, by far, by the production of transverse $W$'s even after 
a cut on forward $W$'s is imposed (Fig.~\ref{eeggww}). 
Another  property is that photons with  like-sign helicities ($J_Z=0$) 
only produce $W$'s with the same helicity. Moreover, at high energy the photons tend to 
transfer their helicities to the $W$'s with the effect that the dominant configurations 
of helicities are 
${\cal M}^{\smx}_{++;++}$ and ${\cal M}^{\smx}_{--;--}$. We have written a general 
helicity amplitude as  
${\cal M}^{\smx}_{\la_1 \la_2;\la_- \la_+}$ with $\la_{1,2}$ the helicities of the 
photons and $\la_{-,+}$ those of the $W^-$ and $W^+$ respectively.
Complete expressions that
specify our conventions are given in the Appendix. 
As a result, if one keeps away 
from the extreme forward region, the dominant helicity states are 
\beqn
\label{smxdom}
{\cal M}^{\smx}_{\pm \pm;\pm \pm} \simeq 4 \pi \alpha \; \frac{8}{\sin^2 \theta}
\eeqn

\noindent which does not depend on the centre-of-mass energy. The $J_Z=2$ are competitive 
only in the very forward direction (with transfer of the helicity of the photon 
to the corresponding $W$). To wit
\beqn
\label{smxdomj2}
{\cal M}^{\smx}_{\la -\la;\la_1 -\la_1} \simeq 4 \pi \alpha \; 
\frac{2 (1+\la \la_1 \cos\theta)^2}{\sin^2 \theta}
\eeqn

Production of longitudinal $W$'s is totally suppressed especially in the $J_Z=0$ channel. 
Moreover the amplitude decreases rapidly with energy. In the $J_Z=2$ configuration 
the amplitude for two longitudinals is almost independent of the scattering angle as 
well as of the centre-of-mass energy. \\
The anomalous contributions present a sharp contrast. First, as one expects with operators 
that describe the Goldstone bosons, the dominating amplitudes correspond to both 
$W$'s being longitudinal. However, in this case it is the $J_Z=0$ amplitude 
which is by far dominating since only the $J_Z=0$ provides the {\em enhancement factor} 
$\gamma$. To wit, keeping only terms linear in $\dkg$, the helicity amplitudes read
\beqn
{\cal M}^{\smx}_{++LL} \sim 
4 \pi \alpha \times \frac{-8}{\gamma \sin^2 \theta}
&\;\;\;\;&
{\cal M}^{\smx}_{+-LL} \sim 4\pi \alpha \times 2 \nonumber \\
{\cal M}^{{\rm ano}}_{++LL} \sim 
4 \pi \alpha \times \gamma \times \Delta \kappa_\gamma
&\;\;\;\;&
{\cal M}^{{\rm ano}}_{+-LL} \sim 
-4 \pi \alpha \times 4 \times \Delta \kappa_\gamma
\eeqn

This contrasting and conflicting behaviour 
between the standard and anomalous contributions in \ggwwt is rather unfortunate. 
As one clearly sees, when one considers the interference between the \sm and the anomalous 
the {\em enhancement factor } $\gamma$ present in the $J_Z=0$ amplitude is offset by the 
{\em reduction factor} in the same amplitude, with the effect that the absolute deviation 
in the total cross section does not benefit from the enhancement factor 
$\gamma=s/M_W^2=s_{\gamma \gamma}/M_W^2$, and therefore 
we would not gain greatly by going to higher energies. In fact 
this deviation is of the same order as in the $J_Z=2$ cross section or that contributed 
by the transverse states. \\ 
\noindent One may be tempted to argue that since the {\em quadratic} terms in the 
anomalous couplings will provide the enhancement factor $\gamma^2$, these quadratic 
contributions could be of importance. However, as confirmed 
by our detailed analysis, these contributions are negligible: the bounds that we have derived 
stem essentially from the linear terms. Moreover, for consistency of the effective 
chiral Lagrangian approach these quadratic terms should not be considered. Indeed their 
effect would be of the same order as the effect of the interference between the \sm amplitude and 
that of the next-to-next-to-leading-order (NNLO) operators. These higher order terms were neglected when we presented the chiral 
Lagrangian.

\subsection{\eewwt}
The situation is quite different in the \eewwt mode. Here one can fully benefit 
from the {\em enhancement factor} even at the level of the total cross section. This also 
means that as we increase the energy one will improve the limits more dramatically 
than in the \gag mode. To make the point more transparent, we limit   ourselves 
to the high energy regime and make the approximation 
 $\sin^2\theta_w\sim 1/4$. 
The helicity amplitudes are denoted in analogy with those in \gag as 
${\cal M}_{\sigma; \tau_-,\tau_+}^{\smx}$ with $\sigma=-$ referring to a left-handed electron. 
With $\theta$ being the angle between the $W^-$ and the electron beam, 
the dominant \sm helicity 
amplitudes which do not decrease as the energy increases are

\beqn
{\cal M}_{-; 0 0}^{\smx} &\sim -4 \pi \alpha&\;\;\;\;    \times  \;\;\;\; \sin\theta\; \left\{ \frac{14}{3} \right\}  \nonumber \\
{\cal M}_{+; 0 0}^{\smx} &\sim -4 \pi \alpha & \;\;\;\;    \times  \;\;\;\; \sin\theta\; \left\{   \frac{2}{3} \right\}\nonumber \\
{\cal M}_{-; \la -\la}^{\smx} &\sim 4 \pi \alpha &
\;\;\;\;    \times  \;\;\;\;
 2 \lambda \frac{\sin\theta (1-\la \cos \theta)}
{1-\cos \theta}\;\;\;\;\;\;\;\;\; \la=\pm
\eeqn

\noi while the dominant helicity amplitudes contributed by the operators of the chiral Lagrangian 
affect predominantly the $W_L W_L$ amplitude
\beqn
\label{eeanowlwl}
{\cal M}_{-; 0 0}^{\rm ano}&\sim  4 \pi \alpha \times &\;\gamma \;\sin\theta\; 
\left\{ l_9+ 4\tilde{l}_1 +\frac{1}{3} r_9 \right\} \nonumber \\
{\cal M}_{+; 0 0}^{\rm ano} &\sim 4 \pi \alpha \times &\; \gamma\;\sin\theta\; 
\left\{ \frac{2}{3} r_9 \right\}
\eeqn

These simple expressions show that 
the enhancement factor $\gamma$ brought about by the anomalous couplings will affect 
the diagonal matrix elements and thus, even at the level of the $WW$ cross section, one will benefit 
from these enhanced couplings. In case of a polarisation with left-handed electrons 
(or with unpolarised beams) all \cpviol conserving couplings will thus be efficiently probed,
whereas  a  right-handed electron polarisation is mostly beneficial 
only in  a model with $L_{9R}$. These expressions also indicate that, with unpolarised beams,
 the bounds on   $L_{9L}$ will be better than those on $L_{9R}$. The additional contribution 
of $L_{9R}$ to the right-handed electron channel will interfere efficiently (with the enhanced 
coupling) only to the diagonal elements, whereas with left-handed electrons this enhancement 
factor can be exploited even for the non-diagonal matrix elements. In this respect 
the special combination 
$\sim l_9+ 4\tilde{l}_1 + r_9/3$ 
is a {\em privileged} direction (in the chiral Lagrangian parameter space), as far as the 
unpolarised \eewwt is concerned since this combination will be by far best constrained. 
The \cviol violating $g_5^Z$ operator is more difficult to probe.  
The latter 
 only contributes to $W_L W_T$ (see Appendix B) with a weaker enhancement factor 
$\sqrt{\gamma}$ which is lost in the interference with the 
corresponding amplitude in the \sm that scales like $1/\sqrt{\gamma}$.
 Another observation is that because $g_5^Z$ does not 
contribute to the {\em privileged} direction, the results of the fit with the two parameters 
$L_{9R,9L}$ will not be dramatically degraded if one fits with the three parameters 
$L_{9R,9L},g_5^Z$. This will not be the case if instead of $g_5^Z$ one considers a 
3-parameter fit with $\slashL_1$. \\

\subsection{Enhancing the sensitivity through the density matrix}
Although the {\em enhancement factor} $\gamma$ is washed out at the level of the \ggwwt cross
section it is possible to bring it out by considering some combinations of the {\em non-diagonal} elements
of the $WW$ density matrix. The latter involve the products of two ({\em different}) 
helicity amplitudes. In order to access these elements one has to analyse 
the distributions provided by the full kinematical variables of the four fermion final state mediated by \ggwwt
and not just the $W$ scattering angle, which is the single variable to rely on at the \ggwwt level. 
 As we have shown in a previous paper we can,  in a first approximation, 
 simulate the  the 4-fermion final state by reverting to the narrow 
width approximation. In this approximation one correlates the \ggwwt helicity amplitudes with the helicity 
amplitudes for $W\ra f_1 \bar f_2$, lumped up in the polarised decay functions $D$. One then arrives at
the 
the five-fold differential cross section 
$\gamma \gamma \ra f_1 \bar{f_2}\; f_3 \bar{f_4}$ which, for definite photon helicities $\la_{1,2}$, 
writes 
\beqn
\label{fullspincorr}
& & 
\mbox{}
\frac{ {\rm d}\sigma(\gamma(\lambda_1)\gamma(\lambda_2) \ra W^+W^-\ra f_1 \bar f_2 f_3 \bar f_4)}
     { {\rm d}\cos \theta \;\;{\rm d}\cos \theta_-^{*} \;\;{\rm d}\phi_-^{*}\;\;
      {\rm d}\cos \theta_+^{*} \;\;{\rm d}\phi_+^{*} }=Br^{f_1 \bar f_2}_W Br^{f_3 \bar f_4}_W 
\frac{\beta}{32\pi s}
\frac{|\vec{p}|}{\sqrt{s}} 
\left( \frac{3}{8 \pi}\right)^2
\hfill \nonumber \\
\lefteqn{ \hspace*{-.5cm}
\sum_{\lambda_- \lambda_+ \lambda'_- \lambda'_+}
 {\cal M}_{\lambda_1,\lambda_2; \lambda_-\lambda_+} (s,\cos \theta)\;  
{\cal M}_{\lambda_1,\lambda_2; \lambda'_-\lambda'_+}^{*} (s,\cos \theta) \;
\;
D_{\lambda_- \lambda'_-} (\theta_-^{*} ,\phi_-^{*}) \; D_{\lambda_+ \lambda'_+} (\pi-\theta_+^{*},
\phi_+^{*}+\pi)
}
\nonumber \\
\lefteqn{ 
\equiv
\frac{{\rm d}\sigma(\gamma(\lambda_1)\gamma(\lambda_2) \ra W^+W^-)}{{\rm d}\cos \theta}
\left( \frac{3}{8 \pi}\right)^2 Br^{f_1 \bar f_2}_W Br^{f_3 \bar f_4}_W
} 
\nonumber \\
\lefteqn{ 
\sum_{\lambda_- \lambda_+ \lambda'_- \lambda'_+}
\rho_{\lambda_- \lambda_+ \lambda'_- \lambda'_+}^{\lambda_1,\lambda_2}\;
D_{\lambda_- \lambda'_-} (\theta_-^{*} ,\phi_-^{*}) \;
D_{\lambda_+ \lambda'_+} (\pi-\theta_+^{*}, \phi_+^{*}+\pi) 
}
\nonumber \\
&{\rm with}& \hspace*{.6cm}\rho_{\lambda_- \lambda_+ \lambda'_- \lambda'_+}^{\lambda_1,\lambda_2}(s, \cos\theta)=
\frac{ {\cal M}_{\lambda_1,\lambda_2; \lambda_-\lambda_+} (s,\cos \theta)\;  
{\cal M}_{\lambda_1,\lambda_2; \lambda'_-\lambda'_+}^{*} (s,\cos \theta)} 
{ \sum_{\lambda_- \lambda_+} |{\cal M}_{\lambda_1,\lambda_2; \lambda_-\lambda_+}(s,\cos \theta)|^2},
\eeqn
\noindent where $\theta$ is the scattering angle of the $W^-$ and $\rho$ is 
the density matrix that 
can be projected out. 

The fermionic tensors that describe the decay of the 
$W$'s are defined 
as in \cite{Fernandeeww}. In particular one expresses 
everything with respect to the $W^-$ where the arguments of the $D$ functions 
refer to the angles of  the particle 
%
%
({\em i.e.} the electron, not the anti-neutrino),
in the rest-frame of the $W^-$, taking as a reference axis 
the direction of flight of the $W^-$ (see~\cite{Fernandeeww}). The $D$-functions to use are therefore
$D^{W^-}_{\la,\la'}(\theta^*, \phi^*)\equiv D_{\la,\la'}$, satisfying 
$D_{\la_1,\la_2}=D_{\la_2,\la_1}^*$ with $\la_i=\pm,0$, and:
\beqn
\label{dfunctions}
D_{+,-}=\frac{1}{2} (1-\cos^2 \theta^*) e^{2i\phi^*},& &
D_{\la,0}=-\frac{1}{\sqrt{2}} (1-\la \cos \theta^*)\sin \theta^* e^{i\la \phi^*}, \\ \nonumber
D_{\la,\la}=\frac{1}{2} (1-\la \cos \theta^*)^2,& &D_{0,0}=\sin^2 \theta^*.
\eeqn
In the decay $W^- \ra e^- \nu_e$, the angle $\theta^*$ is directly related to the energy of the electron (measured in 
the laboratory frame):
\beqn
\label{thetastar}
\cos\theta^*=\frac{1}{\beta} \left( \frac{4 E_e}{\sqrt{s}} -1 \right).
\eeqn

This approximation is a good description of the 4-fermion final state. It also helps
make the enhancement factor in ${\cal M}^{{\rm ano}}_{++LL}$
transparent. Indeed, an inspection of the helicity amplitudes suggests that, in order to maximise 
the effect of the anomalous coupling in \gag, one looks at the interference 
between the above amplitude with the dominant tree-level amplitude, namely 
${\cal M}^{\smx}_{++;++}$. Therefore the elements of the density matrix 
in \gag which are most sensitive to the {\em enhancement} factor $\gamma$ are 
$\rho_{00 \la \la}$ and $\rho_{\la \la 00}$
\beqn
\label{rho00ll}
\rho_{00;\la \la} \sim \frac{8 \times \gamma \times \Delta \kappa_\gamma}{\sin^2 \theta} \sim 
\rho_{\la \la;00}
\eeqn

This particular combination is modulated by the weights introduced by the products of the 
$D$ functions. Of course, averaging over the fermion angles washes out the non-diagonal 
elements. The best is to be able to reconstruct all the decay angles. However, even in 
the best channel corresponding to the semi-leptonic decay there is an ambiguity in assigning 
the correct angle to the correct quark, since it is almost impossible to tag the charge of the jet. 
Therefore the best one can do is to apply an averaging between the two quarks. 
This unfortunately has the effect of
reducing (on average) 
the weight of the $D$-functions. Indeed,  take first the optimal case of the weight ${\cal
W}_{00 \la \la}$
associated to the density matrix elements of interest (Eq.~\ref{rho00ll})
\beqn
\label{optimalweight}
{\cal W}_{00 \la \la}&=&2 Re\left(
D_{0 \lambda} (\theta_-^{*} ,\phi_-^{*}) \;
D_{0 \lambda} (\pi-\theta_+^{*}, \phi_+^{*}+\pi) \;+\; 
D_{\lambda 0} (\theta_-^{*} ,\phi_-^{*}) \;
D_{\lambda 0} (\pi-\theta_+^{*}, \phi_+^{*}+\pi) \right) \nonumber \\
&=& \sin\theta_-^{*}\;\sin\theta_+^{*}\; (1-\la \cos\theta_-^{*})\;(1+\la \cos\theta_+^{*}) 
\cos(\phi_-^{*} - \phi_+^{*})
\eeqn

After averaging  over the quark charges, one has a weight factor with a mean value that is 
reduced by 
$\sim 2.4$:
\beqn
{\cal W}_{00 \la \la}^{sym}=
\sin\theta_-^{*}\;\sin\theta_+^{*}\; (\la- \cos\theta_-^{*})\;\cos\theta_+^{*} 
\cos(\phi_-^{*} - \phi_+^{*})
\eeqn

This is reduced even further if unpolarised photon beams are used since not only 
the $J_Z=2$ 
contribution does not give any enhanced coupling but also the two $J_Z=0$ conspire to give a smaller 
weight than in Eq.~\ref{optimalweight} . Recall that the helicity $\lambda$ in the above tracks the helicity of the photons in 
a (definite) $J_Z=0$ state. Thus one should prefer a 
configuration where photons are in a $J_Z=0$ state.

The above description of the four-fermion final states does not take into account the smearing due 
to the final width of the $W$ and thus one can not implement invariant mass cuts on the decay
products. This is especially annoying since these four-fermion final states 
( as generated through 
the resonant \ggwwt)  can also be generated through other sets of diagrams which do not proceed 
{\it via} \ggwwt. These extra contributions should therefore be considered as a background
\footnote{Note however that some of these extra ``non doubly resonant" contributions also 
involve an anomalous $\Delta \kappa_\gamma$ contribution (single $W$ production diagrams).}.
In a previous investigation\cite{enslapp635} dedicated to these four-fermion final states within the standard model,
we have shown that it was possible to implement a simple overall reduction factor due to smearing 
and invariant mass cuts which when combined with the fully correlated on-shell density matrix 
description reproduces the results of the full calculation based on some 21 diagrams (for the
semi-leptonic channel). Agreement between the improved 
density matrix computation and that based on the full set of diagrams is 
 at the $1-2\%$ level,  if the requirement that very forward electrons 
are rejected is imposed. Since we want to fit the kinematical variables of the electron 
such a cut is essential anyhow. The same overall reduction factor can be implemented even in the 
presence of the anomalous. Even though the $1-2\%$ agreement on the integrated cross sections
 may seem to be  very good, one should also make sure that the same level of agreement is maintained
for the various distributions (see the analysis in \cite{enslapp635}). 
Therefore, we have also analysed the results based on the full set of diagrams contributing to 
$\gamma \gamma \ra l^\mp \bar{\nu}_l jj$ with the inclusion of the anomalous couplings. \\

It is worth pointing out, that the exploitation of the full elements of the density matrix in 
\epemt has been found to improve the results of the fits
\cite{Barklowfits,Coutureee4fano,Moijapan,Gounariserato}
. In \eewwt the greatest improvement 
is expected particularly in multi-parameter fits, since different parameters like $g_5^Z$ and 
$L_9$ affect different helicity amplitudes and thus the use of all the kinematical variables 
of the 4-fermion final states allows to disentangle between these parameters. As we have seen, 
in \ggwwt, in the particular case of the next-to-leading order operators of the effective chiral 
Lagrangian it is impossible to disentangle between the different operators
since they all contribute to the same Lorentz structure. The situation would have been different 
if we had allowed for the couplings $\lambda_\gamma$. In this case, counting rates with 
the total cross section would not differentiate between $\Delta \kappa_\gamma$ and 
$\lambda_\gamma$ but an easy disentangling can be done trough reconstruction of the 
density matrix elements; $\lambda_\gamma$ contributes essentially to the transverse modes 
(see Appendix A). Nevertheless 
in our case the density matrix approach does pick up the important enhancement factors and
therefore,
as we will see, provides more stringent limits than counting rates or fitting on the 
angle of the $W$ alone.  

\section{Limits from \ggwwt}
The best channel where one has least ambiguity in the reconstruction of the kinematical variables of
the four-fermion final states is the semi-leptonic channel. Since $\tau$~'s may not be 
well reconstructed, we  only consider the muon and the 
electron channels. For both the analysis based on the improved narrow 
width approximation\cite{enslapp635} and 
the one based on taking into account all the diagrams, we impose the following set of cuts
on the charged fermions:
\beq
|\cos \theta_{l,j}|<0.98\;\;\;\;\;\;\;\;\;\;\;\;\;\;\;\;\;\;
\cos <l,j><0.9 
\eeq
Moreover we also imposed a cut on the energies of the charged fermions:
\beq
E_f > 0.0125 \sqrt{s}.
\eeq

We take $\alpha=\alpha(0)=1/137$ for the $WW\gamma$ vertex as 
we are dealing with an on-shell photon.
We  take the $W$ to have a mass $M_W=80.22$~GeV. For the computation of the complete 
four-fermion 
final states we implement a $W$ propagator with a fixed width $\Gamma_W=\Gamma_W(M_W^2)=2.08$~GeV.
The same width enters the expression of the reduction factor in the improved narrow 
width approximation that takes into account smearing, 
see\cite{enslapp635}.
The partial width  of the $W$ into jets and $l \bar{\nu_l}$ is 
{\em calculated} by taking at the $W$ vertex the effective couplings 
%
%
$\alpha(M_W^2)=1/128$ and $\sin^2\theta_W=0.23$. 

\begin{table}
\begin{center}
\begin{tabular}{|c|c|c|c|c|c||}\hline
\multicolumn{6}{|c|}{$\sqrt{s} = 400$~GeV   (${\cal L}=20fb^{-1}$)}\\ \hline
Polar &$\sigma$&counting rate&$\cos \theta_{jj}$ &ML& ML \\
$\la_1 \la_2$&(fb)&&&$j\neq\bar{\j}$&$j\leftrightarrow\bar{\j}$ \\ \hline
 unp    &2087  & 3.58 &3.57 & 3.11 &3.50   \\
 unp    & (2064) & (3.59) & (3.58)&(3.12)  &(3.53)   \\
&  & &  & & \\
 $++$     &  2310 &3.60 &3.60 & 2.21 & 3.14   \\
 $++$     &  (2288) &(3.59) &(3.59) & (2.22) &(3.16)    \\
&  & &  & & \\
  $--$    &  2184 &3.76 & 3.76 & 2.26 &3.27\\
  $--$    &  (2186) & (3.77)& (3.77) & (2.27)&(3.29) \\
&  & &  & & \\
  $+-$     & 1926 & 3.47&3.40 & 3.18 &  3.26 \\
  $+-$     & (1893) & (3.49)  & (3.42)& (3.20) &(3.29)   \\
\hline
\hline   
\multicolumn{6}{|c|}{$\sqrt{s} = 800$~GeV  (${\cal L}=80fb^{-1}$)}\\ \hline
 unp    &1154 &2.47  & 2.47&1.20 &  2.28   \\
unp    &(1131) &(2.50)  & (2.49)& (1.21) & (2.32)   \\
&  & &  & & \\
  $++$     & 1290&2.42 &2.41 & .65 &1.40 \\
  $++$     & (1274)& (2.43)& (2.42)& (.66) &(1.41)  \\
&  & &  & & \\
 $--$    &  1258& 2.60&2.60 & .66 & 1.43  \\
$--$    &  (1232)&(2.64) & (2.63)&(.66)  &(1.45)   \\
&  & &  & & \\
 $+-$     & 1034 & 2.43& 2.38& 1.88 & 2.02 \\
 $+-$     & ( 1010) &(2.46) & (2.41)& (1.91) &(2.06)   \\ 
\hline  
\hline
\multicolumn{6}{|c|}{$\sqrt{s} = 1600$~GeV  (${\cal L}=320fb^{-1}$)}\\ \hline
 unp    & 377& 2.08 &2.07 & .33 & 1.35 \\
 unp    & (361)&(2.12)  &(2.09) &(.33)  &(1.37)   \\
&  & &  & & \\
 $++$     & 389 & 2.04& 1.98& .17 & .43\\
 $++$     & (377) & (2.04) &(1.95) &(.17)  &(.43) \\
&  & &  & & \\
 $--$    &  447 & 2.18& 2.10& .17 & .43 \\
 $--$    &  (427) & (2.22)&(2.14) & (.17) & (.43) \\
&  & &  & & \\
 $+-$     & 335 &2.05 &2.01 & 1.11& 1.29\\
 $+-$     & (320) & (2.09)& (2.05)&(1.14)  & (1.32) \\
\hline   
\end{tabular}
\vspace*{-.8cm}
\end{center}
\caption{\label{tablegg}{\em $95\%$CL upper limits on $|L_\gamma|$ obtained from 
the  total cross section measurement (counting rate), a fit to  the reconstructed 
$W^+$ angle ($\cos \theta_{jj}$) as well as from a maximum likelihood fit 
using the full kinematical variables both in the case where the charge of the 
quark is assumed to have been determined ($ j\neq \bar{\j}$) as well as when an averaging 
on the jets has been implemented. The limits are only based on the signature 
$\gamma\gamma\to e^- \bar{\nu}_e u \bar{d}$ calculated by taking into account the full 
set of diagrams. The corresponding results based on the ``improved" narrow width 
approximation are given between parentheses.}}
\end{table}


We will first discuss the results obtained for the specific channel
 $\gamma \gamma  \ra e^- \bar{\nu}_e u \bar{d}$. We will compare the results obtained 
through the approximation based on $\gamma \gamma  \ra W^+ W^- \ra e^- \bar{\nu}_e u \bar{d}$ 
including full spin correlations 
as described in the previous section with those obtained with a simulation which takes into account 
the full set of 4-fermion diagrams. 
In order to compare different methods and make the connection
with  previous analyses, at the level of $\gamma \gamma  \ra W^+ W^-$, which relied only on a 
fit to either the total cross section or the angular distribution
of the $W$'s, we present the results of three different methods for extracting limits on the
anomalous couplings.
The first is a simple comparison between the total number of events 
with the expected standard model rate (``counting rate").
The second is a $\chi^2$ fit on the $\theta_{jj}$ distribution,
$\theta_{jj}$ being the angle of the $jj$ system with the beam pipe,
which corresponds to the angle of the $W$ in the center of mass frame.
Finally, we evaluate the accuracy that a full event-by-event maximum 
likelihood (ML) fit reaches.  The latter analysis exploits 
the 5 independent variables 
describing the kinematics:
the polar and azimuthal angles of the $jj$ and $l\nu$ pairs
in the frame of the decaying ``$W$'s'' and the polar angle
of the ``$W$'' pairs in the center-of-mass of the colliding beams. \\

Let us be more specific about how we have exploited 
the (extended) maximum likelihood method both in \epemt and \gag. The anomalous couplings 
$(L_{9L},L_{9R},...)$ represent the components of a vector $\vec{p}$.
The (fully) differential cross section defines  a probability
density function.  Given $f(\vec{x};\vec{p}) d\vec{x}$, the average number of 
events to be found at a phase space point $\vec{x}$ within
$d\vec{x}$,
we calculate the likelihood function
(${\cal L}$) using a set of $N$ events \cite{Barlow}:
\beq
\ln{\cal L}=\sum^{N}_{1} \ln f(\vec{x}_i;\vec{p}) - n(\vec{p}).
\eeq
with $n(\vec{p})=\int f(\vec{x};\vec{p}) d\vec{x}$, the theoretical 
total number of events expected.
For a given set of experimental measurements $x_i$, 
${\cal L}$ is a function of the parameters $\vec{p}$ we would like to 
determine.  The best estimate for $\vec{p}$  is the one that maximises
the likelihood function $\cal L$ or, equivalently, $\ln\cal L$.  
The statistical error\footnote{We have not made any effort to include systematic errors 
in our analysis.} in the estimation can be
easily measured as ${\cal L}$ exhibits a Gaussian behaviour around the 
solution.  However, it is not necessary to reproduce realistic data
to know how well the parameters can be determined.
For a large number of events, the statistical error on the set of 
parameters $\vec{p}$ can be evaluated simply by
\beq
\label{errorml}
\overline{(p_i-\overline{p}_i)
(p_j-\overline{p}_j)}=\left[\int \frac{1}{f(\vec{x};\vec{p})}\left(
\frac{\partial f(\vec{x};\vec{p})}{\partial p_i}
\frac{\partial f(\vec{x};\vec{p})}{\partial p_j} \right) d\vec{x}\right]^{-1},
\eeq
which is easily computed numerically.  With more than one parameter, 
the right-hand side of Eq.~\ref{errorml} is understood as a matrix
inversion.  \\

From the qualitative
arguments we have given as regards the effect of polarisation of the photons, we study all possible 
combinations of circular polarisation of the photons as well as the case of no polarisation. 
At the same time, having in view the efficient reconstruction of the non-diagonal elements of 
the density matrix we consider the case of being able to identify the charge of the jet. 
For the  analysis conducted with the full set of diagrams,  we allow the 
invariant masses of the jet system (and the leptonic system) to be extra kinematical parameters in the fit. 
No invariant mass cuts have been implemented so that to exploit the full statistics. \\
Our results are assembled in Table~\ref{tablegg}. Note that, as we explained in a previous paper
\cite{enslapp635}, one should not
expect the cross sections for the two $J_Z=0$ ($++$ and $--$) to be equal, because of the chiral 
structure of the lepton-W coupling and our choice of cuts (none on the neutrino). This is the reason
one should be careful when combining the results of the charged conjugate channel (with $e^+$). In
fact, if a $++$ setting is chosen for the \gag collider, then the corresponding results for the
channel $\gamma \gamma  \ra W^+ W^- \ra e^+ \nu_e  \bar{u} d$ should be read from the 
$--$ entry in Table~\ref{tablegg}. \\
This table gives the $95\%$CL limits obtained on $|L_\gamma \equiv L_{9L}+L_{9R}+4 \slashL_1|$ 
from  different fitting 
methods.
We have considered three centre-of-mass energies for \gag collisions: $400$, $800$ and $1600$GeV.
These correspond to $80\%$ of the energy of the \epemt collider. We also assumed a fixed photon
energy and thus no spectrum.   
The luminosity is assumed to be ${\cal L}_0=20 (.4/{\sqrt s}(TeV))^2$. \\

The first important conclusion is that {\em irrespective} of the method chosen to extract the limits
and for {\em all} centre-of-mass 
energies, the limits one extracts from an analysis based on the full set of
diagrams and those based on the density matrix approximation are, to a very good precision,
essentially the same. The errors on the limits are within $2\%$. Another conclusion which 
applies to all energies relates to the limit one extracts from fitting only the $W^+$ scattering 
angle, that is, from an \ggwwt analysis. One gains very little compared to a limit extracted from 
a counting rate. Fortunately, the information contained in the full helicity structure 
(fitting through a ML with all kinematical variables) is quite essential. The bounds improve 
sensibly in this case, especially so when the energy increases and if one selects a $J_Z=0$ 
setting.\\
\noindent If one restricts the analysis to fits on the $W$ scattering angle only 
($\cos\theta_{jj}$),  or to bounds 
extracted from a simple 
counting rate, the improvement one gains as the energy increases is very modest. In fact this modest
improvement is due essentially to the slightly larger statistics that we obtain at these higher 
energies. These larger statistics have to do with the fact that the assumed luminosity more than 
make up for the decrease in the cross sections. We have shown in the previous section how this comes
about and why it is essential to recover the {\em enhancement} factor $\gamma$ in the $J_Z=0$ amplitude 
by reconstructing the elements of the density matrix. Indeed as our results show, polarisation 
(with a $J_Z=0$ setting) is 
beneficial only when combined with a ML fitting procedure. A most dramatic example that shows 
the advantage of this procedure is found at $1600$GeV where the improvement over the counting rate 
method is more than an order of magnitude better in the case of recognising the jet charges. Note
that our results in this case, when comparing between the three energies, 
do reflect the factor $\gamma$ enhancement. On the other hand in the $J_Z=2$ polarisation 
with a ML brings only about a factor 2 improvement. At high energy ($\geq 800$GeV) the tables 
also confirm the reduction ($\sim 2.4$ that we discussed above) when a symmetrization 
($j\leftrightarrow{\bar{\j}}$) in the two 
jets is carried out. Moreover,  as expected, we find that when this symmetrization is performed 
the results with unpolarised beams are much worse than {\em any} of the $++,--,+-$ settings 
(see Eq.~3.9-3.10). Therefore it clearly pays to have polarisation, choose $J_Z=0$ and perform 
a maximum likelihood method. One undertone though is that at 400GeV one still can not fully 
exploit the {\em enhancement} factor and consequently polarisation and maximum likelihood 
fare only slightly better than an unpolarised counting rate. Nonetheless, already at this 
modest energy, with $20fb^{-1}$ of integrated luminosity and with only the channel 
$e^- \bar{\nu}_e u \bar{d}$ one can put  the bound $L_\gamma \sim 3$. At $1.6$TeV 
with a $++$ setting one can reach $.43$ after including an averaging on the jet charges. 
Including all semi-leptonic channels one attains $L_\gamma \sim .14$. These limits are thus of the same order 
as those one has reached on the parameter $L_{10}$ for example, from present high precision 
measurements. 

\section{Comparing the results of detailed fits in \epemt and \gag}
As the results of the previous analysis show, the \gag collider places excellent bounds 
on the anomalous $WW\gamma$ coupling. However one obvious disadvantage is that \ggwwt can not 
disentangle between different operators of the chiral Lagrangian and therefore between the indirect 
effects of different models of symmetry breaking. 
Since, as may be seen through the helicity amplitudes of 
\eewwt (Eqs.~\ref{eeanowlwl},~\ref{eewwanoapp}), the different operators 
of the chiral 
Lagrangian have ``different signatures" in the \epemt mode, one should be able to disentangle 
between different operators or at least give bounds on all of them in \epemt, and not just probe
{\em one} specific 
combination of them as in \gag. Therefore, as far as the anomalous couplings are concerned, one 
should question
whether it is worth supplementing the next linear collider with a \gag option. To answer this, one
needs to know whether the limits one gets from the \epemt mode are as good, or at least 
competitive, with those one extracts from the \gag mode. Indeed, 
it is already clear from our qualitative arguments concerning \eewwt, that though the chiral 
Lagrangian operators affect the various helicity amplitudes in a discernible way, the greatest 
sensitivity (involving the {\em enhanced couplings}) stems from one particular helicity 
amplitude that involves a specific combination of the chiral Lagrangian operators. As a result 
one should expect that if one conducts an analysis in \eewwt to scan the entire parameter 
space of the anomalous operators, one would not get stringent limits on all the parameters but 
expect that one particular combination of parameters to be much better constrained than other 
directions in the space of anomalous parameters. If the
bounds on the latter are too loose they may not be useful enough to test any model, in the sense that 
models of symmetry breaking 
predict smaller values. On the other hand, by combining these bounds with the very stringent 
limits derived from \gag one may be able to reach a better level of sensitivity.  In the following 
we will attempt to address these points. We will compare the results of \ggwwt and \eewwt 
in the case where one has imposed the global custodial symmetry, which in effect allows 
only two parameters ($L_{9L}$ and $L_{9R}$) and see how the \eewwt channel fares when we include the 
extra parameters 
$L_c$ and $\slashL_1$. \\

Various analyses\cite{wwanolep2,Coutureee4fano} including complete calculations of the four-fermion 
final states in \epemt and exploiting the ML techniques have  been 
conducted recently\footnote{Another very recent analysis\cite{Gounariserato} is based on the technique of 
the optimal observables.}
. We differ from these by our choice of 
anomalous couplings. 
We allow in particular for the \cviol violating $g_5^Z$ parameter as well 
as the custodial symmetry breaking parameter $\slashL_1$. Moreover 
we found it important to conduct our own analysis for \epemt in order to 
compare, on the same footing, the results of the \gag and \epemt analyses. 
We will only take into account the semi-leptonic final states. In this 
comparison we show the results based on the complete set of 
4-fermion semi-leptonic final state including the special case of an 
$e^\pm$ in the final state which for $\epem \ra 4f$  involves a larger set of 
diagrams. In the present analysis we only consider the case of 
unpolarised 
electron beams. 
The benefits of beam polarisation and how the luminosity 
in \epemt could be most efficiently shared between  the two  electron helicities will 
be studied in a forthcoming publication\cite{Marcpapieranomal}.

First of all, our detailed ML fit of \eewwt does confirm that for all energies ($500,1000,2000$GeV) 
there is a privileged direction involving a specific combination of 
$L_{9L}, L_{9R}$ and $\slashL_1$ that is far better constrained than any other 
combination. 
This particular combination, $\sim L_{9L} + 4\slashL_{1} + 0.44 L_{9R}$,
 is different from the one probed in \ggwwt and can in fact be deduced 
from our approximate formulae for the dominant
anomalous \eewwt helicity amplitude
(Eqs.~\ref{eeanowlwl},~\ref{eewwanoapp}) that  corresponds to $W_L W_L$ production. 
This combination as extracted from the fit is to be compared with  the combination that appears 
in our approximate formulae for $W_L W_L$ with a left-handed electron 
$\sim L_{9L} + 4\slashL_{1} +  L_{9R}/3$. For a better agreement one notes that one should add 
the contribution of the right-handed electron to which contributes essentially only
$L_{9R}$. This particular behaviour is reflected in our figures that show the 
multi-parameter bounds in the form a pancake. \\
\begin{figure*}[htbp]
\begin{center}
\vspace*{-2.cm}
\hspace*{-5.5cm}\mbox{\epsfig{file=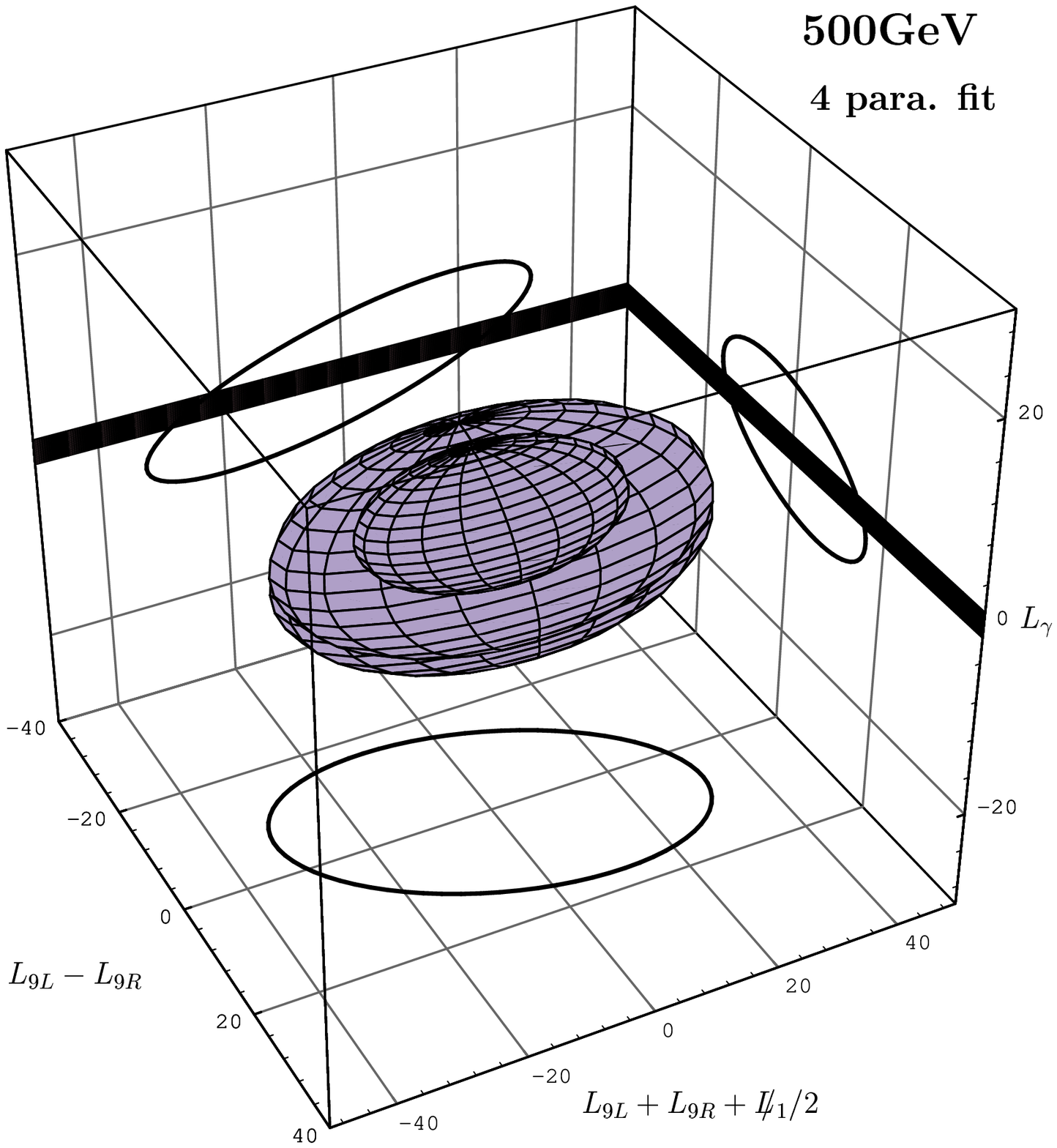,width=10cm}}
\vskip 0.8cm
\hspace*{7cm}\mbox{\epsfig{file=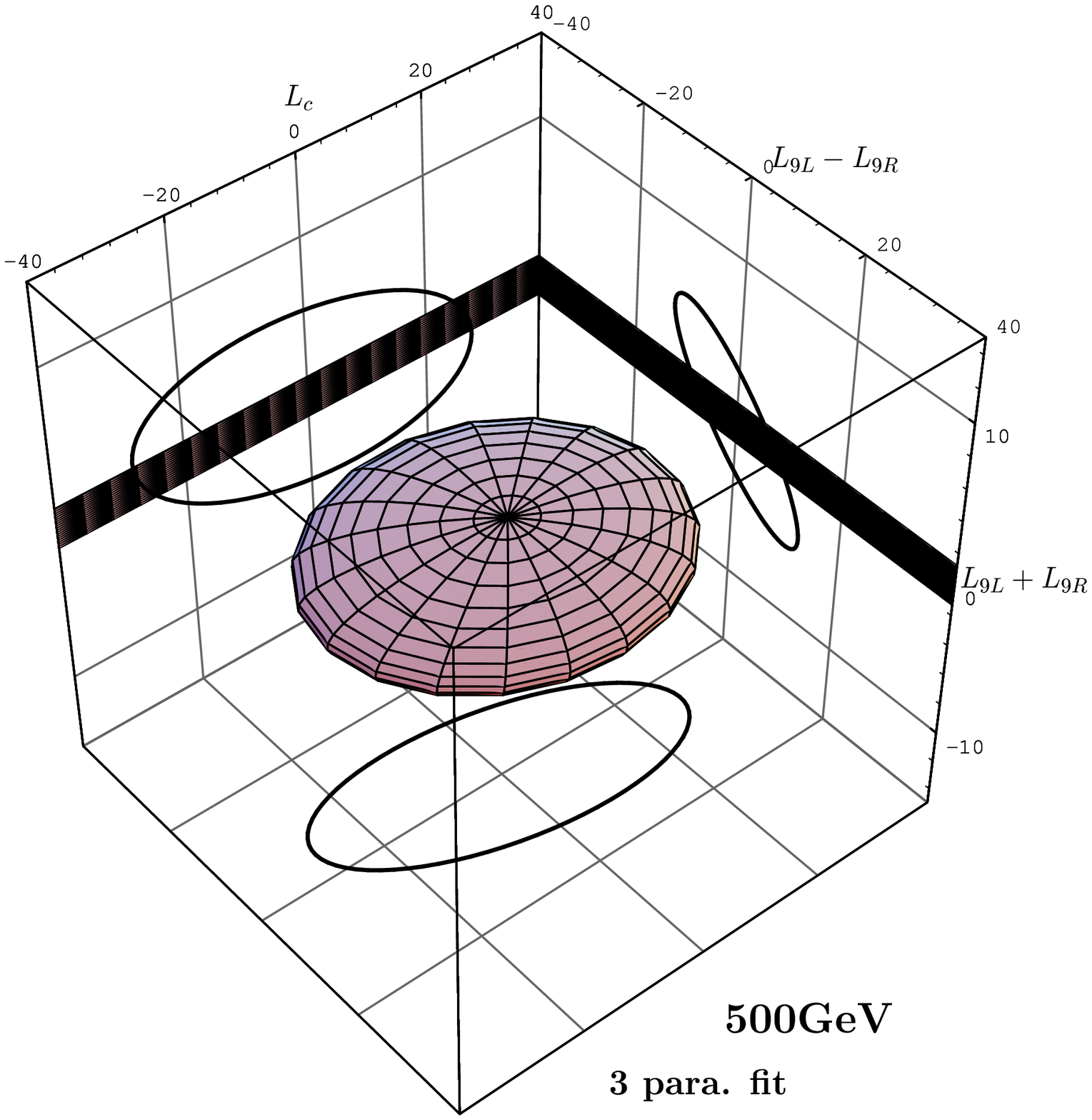,width=8.2cm}}
\vspace*{-.3cm}
\caption{\label{fig5003d4p}{\em The tri-dimensional bounds in the case of 
4 parameters for a fit at $\sqrt{s}=500$GeV ( $\sqrt{s_{\gamma \gamma}}=400$GeV). 
The ellipsoide represents the \epemt bound with the ellipses being the projections 
in the different planes. The limit from \gag with unpolarised photons 
consist of two wafers (planes) whose 
projections on the planes ($L_\gamma$-$L_V$) and ($L_\gamma$-$L_{9L}+L_{9R}-\slashL_1/2$) are 
shown. The two `` \gag wafers" should be visualised as cutting through the ellipsoid. The lower 
figure is the result of a 3-parameter fit with $\slashL_1=0$. 
All results are at $95\%$CL.}}
\end{center}
\end{figure*}
\noi In the case where we allow a global $ SU(2)$ breaking with $\slashL_1 \neq 0$, we have preferred 
to visualise our results by using the set of independent variables 
$L_\gamma=L_{9L} + L_{9R} +4\slashL_{1}$ (representing the \gag direction), 
$L_V= L_{9L} -L_{9R}$ (this would be zero in a vectorial model on the mould of a 
scaled up QCD 
\footnote{ But then a scaled up version of QCD has $L_{10}$ of the same order as $L_{9L}$.}) 
and the orthogonal combination ($\sim L_{9L} + L_{9R} -\slashL_{1}/2$). \\
\begin{figure*}[htbp]
\begin{center}
\mbox{\epsfig{file=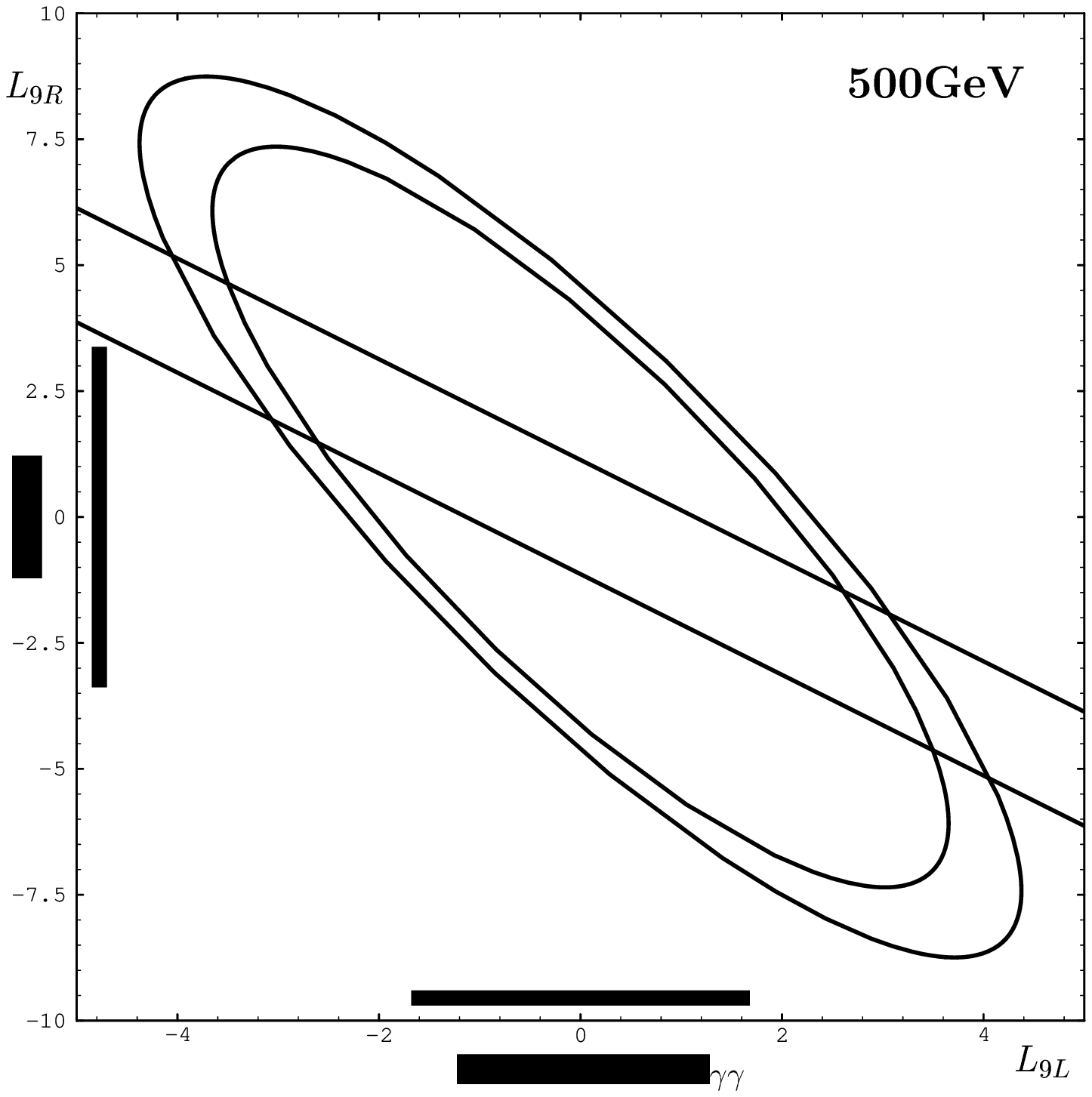,width=13cm}}
\caption{\label{fig5002d}{\em a 2D comparison between \ggwwt and \eewwt at 500GeV. 
The ellipses are the results from \eewwt, with the smaller ellipse  
corresponding to a fit with two parameters ($L_{9L},L_{9R}$) while the larger  one 
is for the case of three parameters ($L_{9L},L_{9R},L_c$). The diagonal band is from 
\ggwwt (unpolarised photons).  
 The ``bars" along the axes 
are the one-parameter fits. The thinner ones (inside the box) are for \epemt and
 the thicker ones (outside the box) are for \gag. 
All results are at $95\%$CL.}}
\end{center}
\end{figure*}
For \epemt at $500$GeV and with  a full 4-parameter 
fit, one sees (Fig.\ref{fig5003d4p}) that 
 the limits  from \eewwt lead to relatively loose bounds that are not competitive with what one
obtains from the \gag mode.  In fact the  parameter space  allows couplings of ${\cal{O}} (10)$ 
and hence it is doubtful that such analysis will usefully probe symmetry breaking.  
Even upon switching off the SU(2) violating coupling $\slashL_1$, the 
multi-parameter bound (Fig.\ref{fig5003d4p}) obtained from \eewwt does not compare well with 
the stringent bounds that one is able to reach in \ggwwt. Even though, in this case 
\ggwwt is blind to $L_c$, the bound on $L_c$ from \epemt is not strong enough to be useful
($|L_c|< 20$). At this energy the benefits of \ggwwt are very desirable, since when combined with 
the limits from \eewwt the parameter space shrinks considerably, even if with little effect 
on the limit on $L_c$. At this energy even in the case of maintaining an exact global SU(2) 
symmetry with only the parameters ($L_{9L},L_{9R}$) remaining, the $L_9$ bound is sensibly reduced 
if one takes advantage of the \gag mode(Fig.\ref{fig5002d}). Note however that the limits from a 
maximum likelihood fit in \eewwt, with an integrated luminosity of  $\int{\cal L}=20fb^{-1}$ 
lead to $|L_{9R}| <7\;\;-\;|L_{9L}| <4$; with a slight degradation if one had included $L_c$ into the
fit. 
As can be seen from  Eq.~\ref{eeanowlwl}, $L_c$ does not 
contribute to the sensitive $W_L W_L$ direction which 
benefits the most from the enhanced coupling. \\

What about if only one parameter were present?
In this case both \eewwt and \ggwwt give excellent limits as  in 
Fig.~\ref{fig5002d}. \gag give slightly better limits, especially in the case of 
$L_{9R}$ where we gain a factor of two in \gag. Note however that this result is 
obtained without the inclusion of the photon spectra. The latter affects much more the 
effective \gag luminosity than the \epemt. If one includes a luminosity reduction factor 
of 10 in the comparison, the results of single-parameter fits would be essentially the same in the 
two modes.  At 500GeV, polarisation in \gag has almost no effect for this physics. It should 
however be  kept in mind that in \eewwt right-handed polarisation would be welcome in fits including 
$L_{9R}$\cite{Marcpapieranomal,Barklowfits,Coutureee4fano}. \\

\begin{figure*}[htbp]
\begin{center}
\vspace*{-3.cm}
\hspace*{-5.5cm}\mbox{\epsfig{file=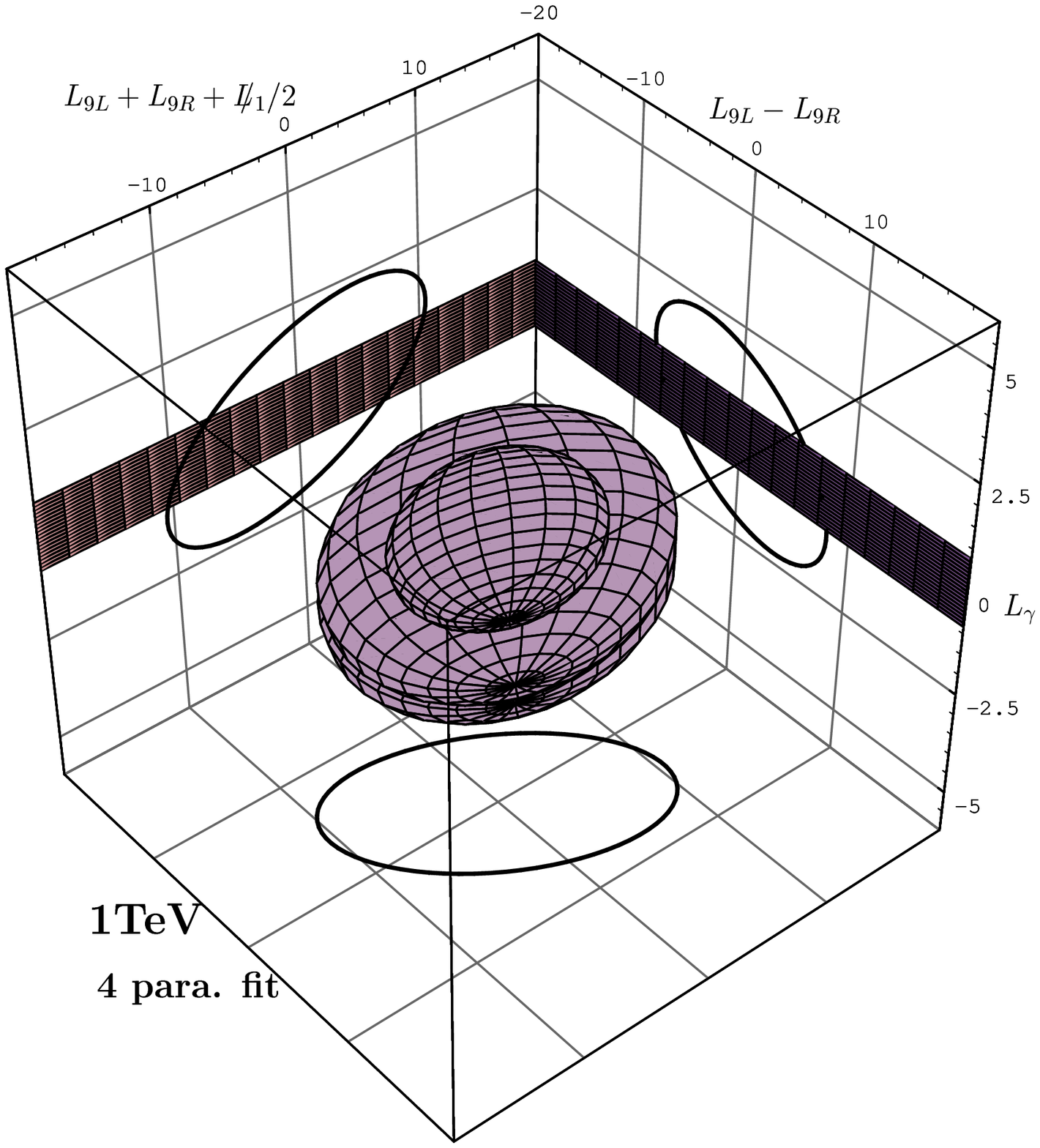,width=10cm}}
\vskip 1cm
\hspace*{5.5cm}\mbox{\epsfig{file=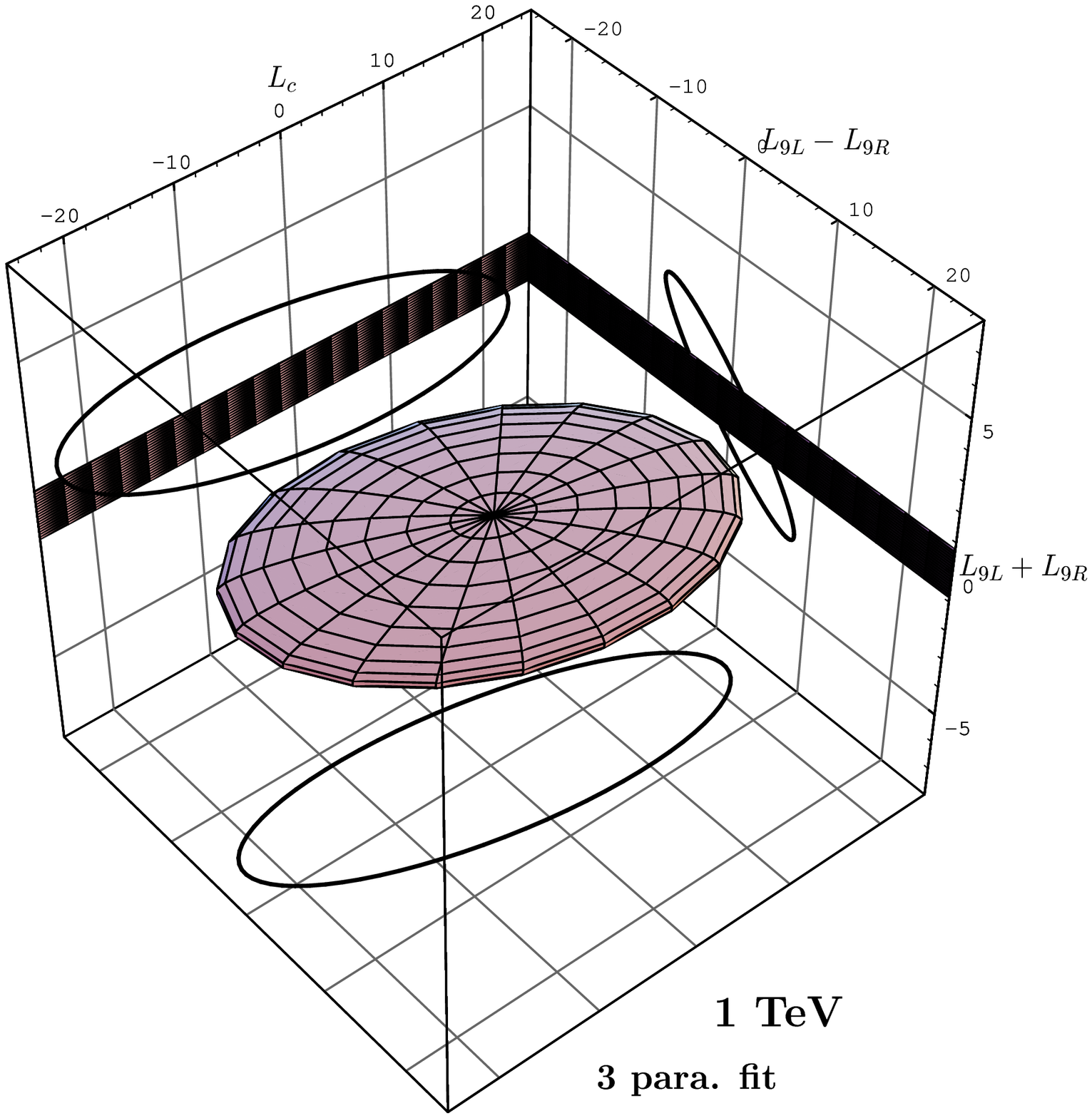,width=8.2cm}}
\vspace*{1cm}
\caption{\label{fig13d4p}{\em As in fig.~\ref{fig5003d4p} but for $\sqrt{s}=1$TeV.}}
\end{center}
\end{figure*}

\begin{figure*}[htbp]
\begin{center}
\vspace*{-3.cm}
\mbox{\epsfig{file=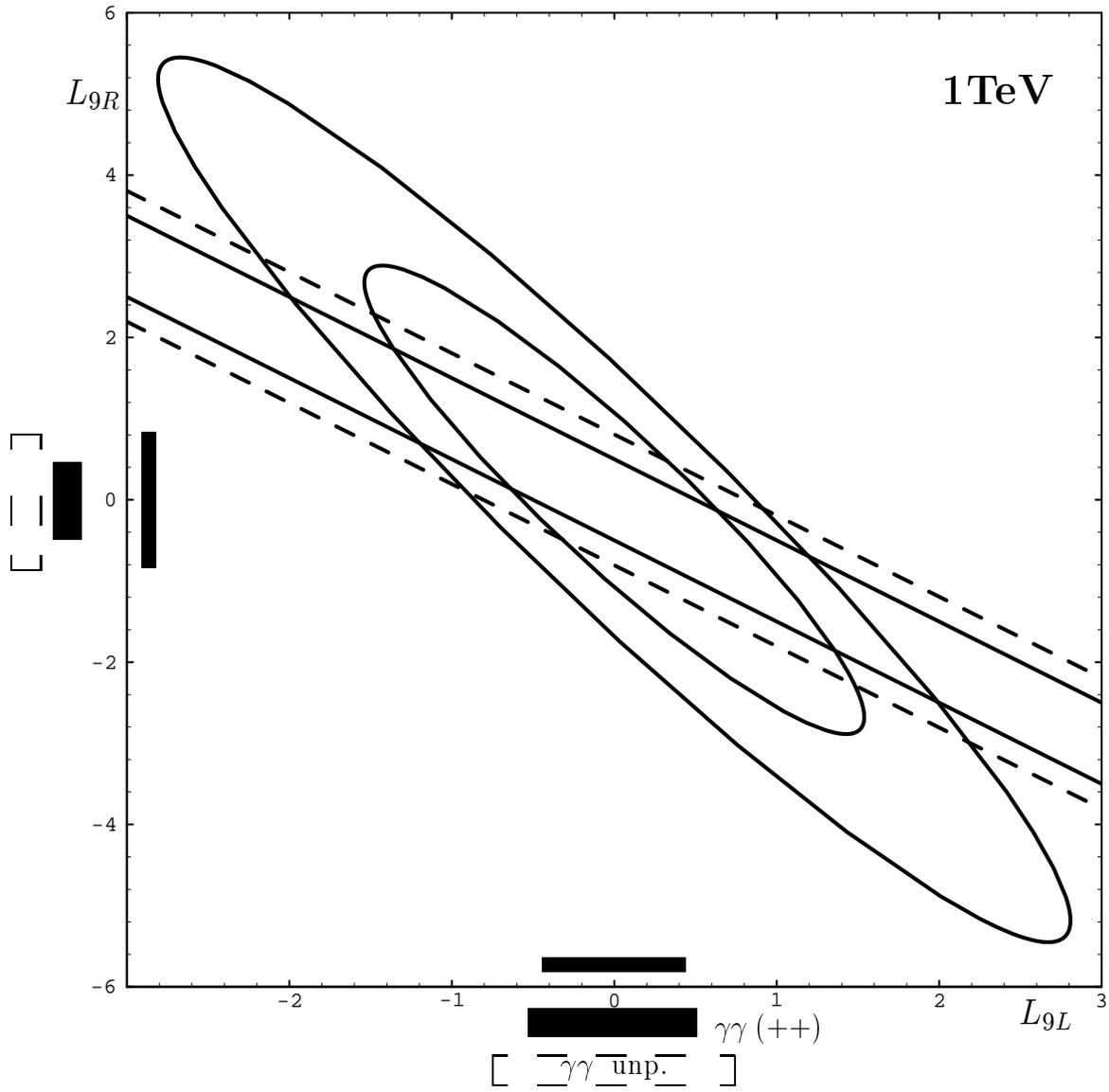,width=15cm}}
\vspace*{1cm}
\caption{\label{fig12d}{\em As fig.~\ref{fig5002d}. 
but with the dashed bands representing the unpolarised result in 
\gag. The additional bands are 
for the case  $J_Z=0$ ($++$). }}
\end{center}
\end{figure*}

\begin{figure*}[htbp]
\begin{center}
\vspace*{-3.cm}
\hspace*{-5.5cm}\mbox{\epsfig{file=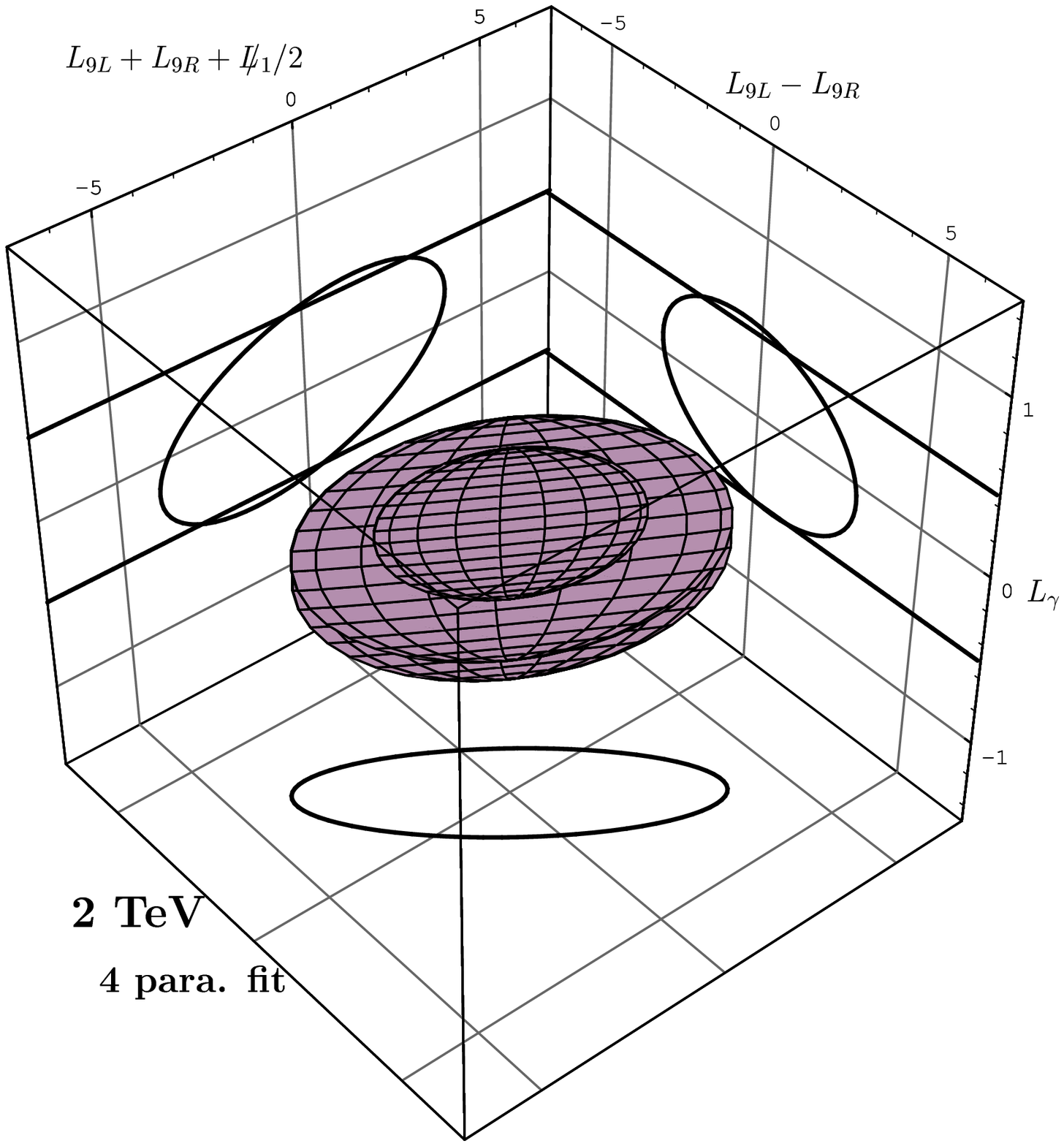,width=10cm}}
\vskip 1cm
\hspace*{5.5cm}\mbox{\epsfig{file=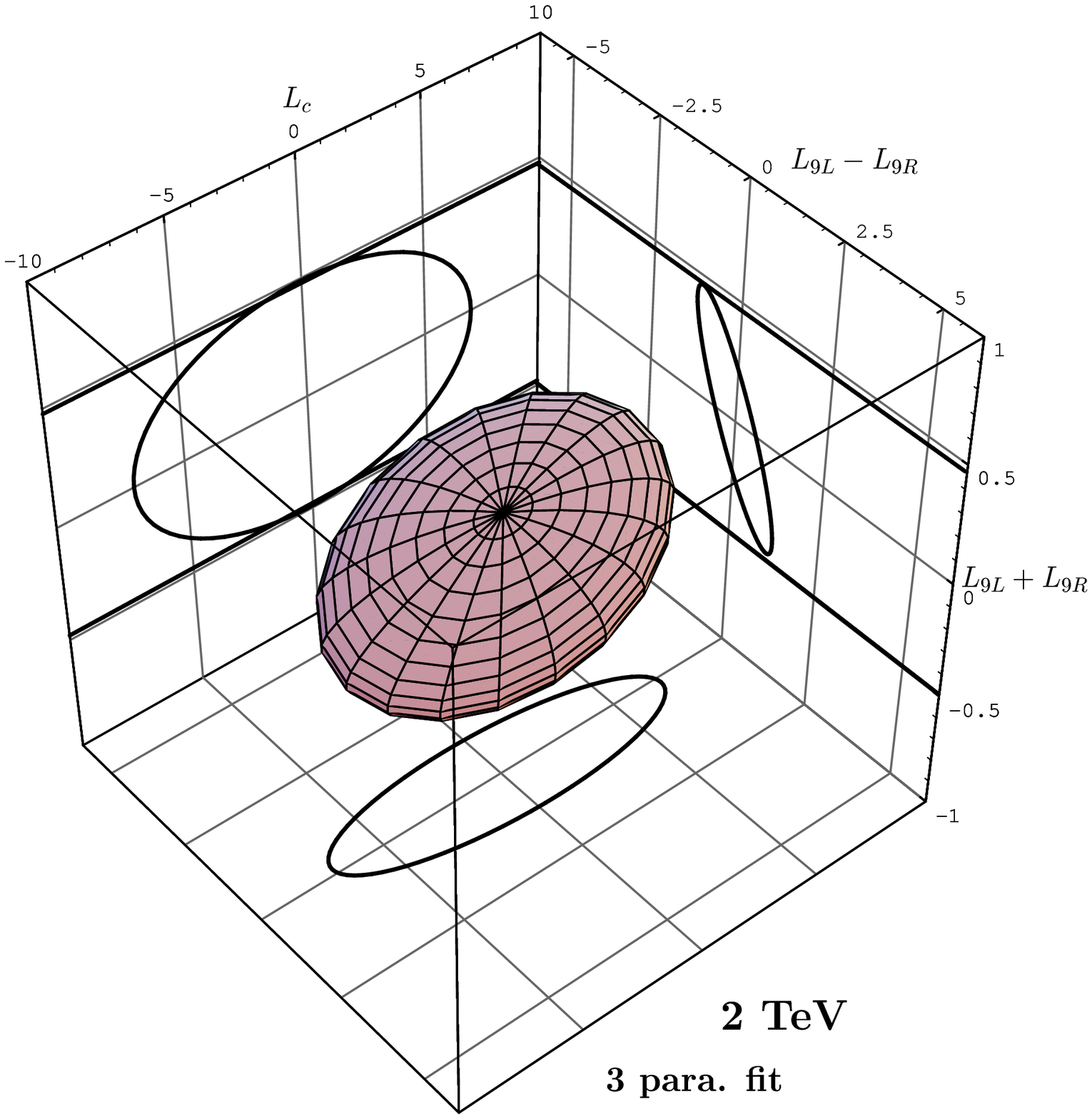,width=8.2cm}}
\vspace*{1cm}
\caption{\label{fig23d4p}{\em As in fig.~\ref{fig5003d4p} but for $\sqrt{s}=2$TeV.}}
\end{center}
\end{figure*}

\begin{figure*}[htbp]
\begin{center}
\vspace*{-3.cm}
\mbox{\epsfig{file=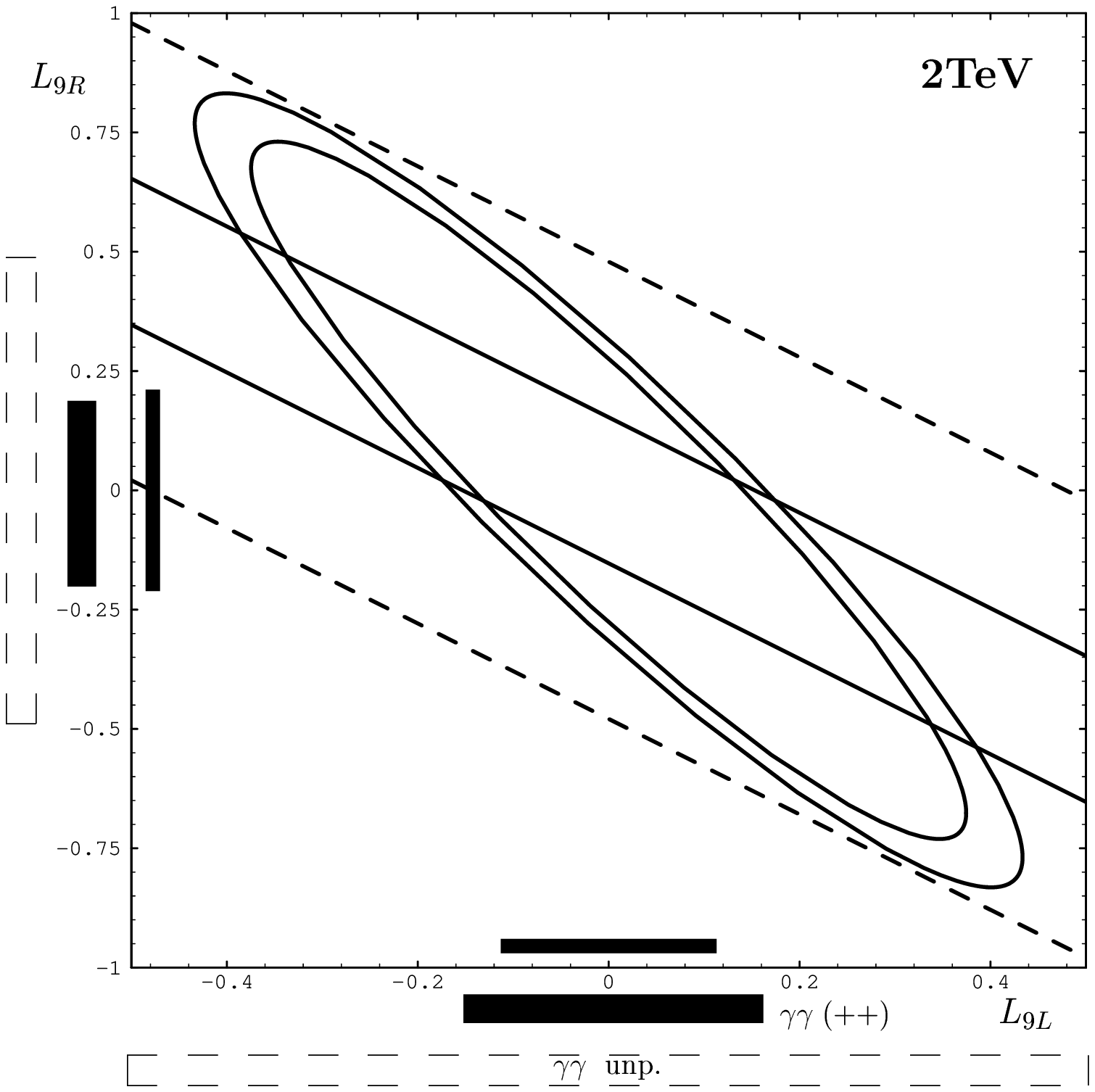,width=15cm}}
\vspace*{1cm}
\caption{\label{fig22d}{\em As fig.~\ref{fig12d}}}
\end{center}
\end{figure*}

As the energy increases, the role of polarisation in \gag becomes important, as we detailed in 
the previous section (see Table~\ref{tablegg}). However the benefits of \ggwwt 
turn out to be  rather mitigated, in the sense that combining the results from \ggwwt 
to those obtained in \eewwt does not considerably reduce the bounds one deduces from \eewwt alone, 
see Figs.~\ref{fig13d4p}-~\ref{fig22d}. 
This is especially true at 2TeV, where in the case of a 4-parameter fit, the \gag limits 
reduce the bounds slightly only if the \gag are in the correct polarisation setting ($J_Z=0$)
(see Figs.~\ref{fig23d4p}-\ref{fig22d}). At 2TeV even if one allows a three parameter fit 
with $L_c$ ($L_9$), an unpolarised \gag collider does not bring any further constraint and one 
gains only if one combines polarised \gag beams with ML methods (Fig.~\ref{fig22d}). For the case of  a one-parameter fit 
our results indicate, that if one has the same luminosity in the \epemt and \gag modes than 
there is practically no need for \ggwwt even when the photon beams are 
polarised (Fig.~\ref{fig22d}).

\section{Conclusions}
We have critically analysed  the usefulness of the \gag mode of the 
next linear collider in probing models of symmetry breaking through 
the effects of anomalous couplings in the reaction \ggwwt. To take full 
advantage of all the information provided by the different helicity 
amplitudes we have taken into account all the contributions to the full 
four-fermion final states and studied the approximation based on the 
``resonant" $W^+ W^-$ final state with complete spin correlations. One of our 
results is that exploiting the full information provided by the 
kinematical variables of the 4-fermion variables, as made possible through a 
fit based on the maximum likelihood method, not only does one obtain 
excellent limits on the anomalous couplings but we improve considerably 
on the limit extracted at the level of \ggwwt. Especially as the 
energy increases (beyond $500$GeV) these limits are further improved  if 
use is made of polarising the photon beams in a $J_Z=0$ setting. One limitation of the 
\gag mode is that within our effective Lagrangian, \ggwwt only probes one collective 
combination of operators, that contribute to the magnetic moment of the $W$ (usually refered 
to as $\dkg$). Disentangling between different operators that could point to different 
mechanisms of symmetry breaking is therefore not possible. We have therefore addressed the 
question of how this information compares to what we may learn from the normal mode 
of the linear collider, \epemt, and whether combining the results of the two modes 
further constrains the models. In order to conduct this comparison we have relied on 
the complete calculation of  the full 4-fermion final state in \epemt and used the same 
analysis (based on the maximum likelihood method) as the one in \gag. It turns out that 
up to 1TeV and in case we allow for more than one anomalous coupling, there is some  
benefit (especially at 500GeV) in having a \gag mode for this type of physics. However 
for all energies, if one only considers one anomalous coupling, there is very little or 
no improvement brought about by the \gag mode over the \epemt mode. Considering that our 
analysis has not taken into account the folding with the luminosity functions which 
will lead to a reduced effective luminosity in the \gag mode, there  seems that for one 
parameter fit there is no need for a \gag mode. At much higher energies (2 TeV) this conclusion 
holds even for multi-parameter fits. 

\newpage
\setcounter{equation}{0}
\def\thequation{\thesection.\arabic{equation}}
\setcounter{section}{0} 
\setcounter{subsection}{0}
\def\thesection{\Alph{section}} 
\def\thesubsection{\thesection.\arabic{subsection}} 
{\Large \bf Appendix}
\section{Helicity amplitudes for \ggwwt} 
\setcounter{secnumdepth}{2} 
\setcounter{equation}{0}
\def\thesubsection{\thesection.\arabic{subsection}} 
\def\theequation{\thesection.\arabic{equation}} 
\subsection{Tree-level helicity amplitudes for $\gam \ra 
W^+W^-$ in the \sm} 
To understand the characteristics of the $WW$ cross-section it is best to 
give all the helicity amplitudes that contain a maximum of information on the 
reaction.
It is important to specify our conventions. We work in the centre of mass 
of the incoming photons and refrain from making explicit the azimuthal dependence 
of the initial state. The total energy is $\sqrt{s}$. 
We take the photon with helicity $\la_1$ ($\la_2$) to be in the $+z$ ($-z$) direction 
and the outgoing 
$W^-$ ($W^+$) with helicity $\la_-$ ($\la_+$)and  4-momentum $p_-$  $p_+$:
\beq
p_\mp^\mu=\frac{\sqrt{s}}{2}(1, \pm \beta \sin \theta,0, \pm\beta \cos \theta)
\eeq 
In the following all $\la_i=\pm$. 
The polarisations for the helicity basis are defined as 

\beqn
\epsilon_1^\mu(\la_1)=\frac{1}{\sqrt{2}} (0,-\la_1,-i,0) \;\;\;\;\;\;\;\;\;& &
\epsilon_2^\mu(\la_2)=\frac{1}{\sqrt{2}} (0,\la_2,-i,0) \\ \nonumber 
\epsilon_3^\mu(\la_-)^*=\frac{1}{\sqrt{2}} 
(0,-\la_- \cos \theta,i,\la_- \sin \theta)& &
\epsilon_4^\mu(\la_+)^*=\frac{1}{\sqrt{2}} 
(0,\la_+ \cos \theta,i,-\la_+ \sin \theta)\\ \nonumber 
\epsilon_3^\mu(0)^*=\frac{\sqrt{s}}{2M_W} (\beta, \sin \theta,0,\cos \theta)
\;\;\;\;\;\;\;\;\;& &
\epsilon_4^\mu(0)^*=\frac{\sqrt{s}}{2M_W} (\beta, -\sin \theta,0,-\cos \theta)
\eeqn

We obtain for the tree-level \sm helicity amplitudes:
\beq
{\cal M}_{\la_1 \la_2; \la_- \la_+} = \frac{4 \pi \alpha}
{1-\beta^2 \cos^2 \theta} \;\;{\cal N}_{\la_1 \la_2; \la_- \la_+} \;\;\; ; \;
\;\; \beta=\sqrt{1-4/\gamma} \; ; \; \gamma=s/\mww
\label{eq:mn}
\eeq
\noindent
where
\beqn
{\cal N}_{\la_1 \la_2; 0 0} &= &\;-\;\frac{1}{\gamma}
\left\{ 
-4 (1+\la_1 \la_2) + (1-\la_1 \la_2) (4+\gamma) \sin^2 \theta  \right\}
\nonumber \\
{\cal N}_{\la_1 \la_2; \la_- 0} &=& \sqrt{\frac{8}{ \gamma}} \;\; (\la_1-\la_2) 
         (1+\la_1 \la_- \cos \theta) \sin \theta  \;
\nonumber \\
{\cal N}_{\la_1 \la_2; 0,\la_+ } &=&\;-\;\sqrt{\frac{8}{ \gamma}} \;\; (\la_1-\la_2) 
         (1-\la_1 \la_+ \cos \theta) \sin \theta  \;
\nonumber \\
{\cal N}_{\la_1 \la_2; \la_- \la_+}&=& \beta (\la_1+\la_2) (\la_-+\la_+) + 
\frac{1}{2 \gamma}\left\{
-8 \la_1 \la_2 (1+\la_- \la_+) +\gamma (1+\la_1 \la_2 \la_- \la_+) 
(3+\la_1 \la_2) \right.  \nonumber \\
&+& \left.  2 \gamma (\la_1-\la_2)(\la_--\la_+)\cos\theta
- 4 (1-\la_1 \la_2)(1+\la_- \la_+) \cos^2\theta \right. \nonumber \\
&+& \left. \gamma (1-\la_1 \la_2)(1-\la_- \la_+) \cos^2\theta \right\}
\label{eq:lesbons} 
\eeqn

With the conventions for the polarisations, the fermionic tensors are defined 
as in \cite{Fernandeeww}. In particular one expresses 
everything with respect to the $W^-$ where the arguments of the $D$ functions 
refer to the angles of  the particle (electron 
not anti-neutrino), in the rest-frame of the $W^-$. The D-functions to use are therefore
$D^{W^-}_{\la,\la'}(\theta^*, \phi^*)\equiv D_{\la,\la'}$, satisfying 
$D_{\tau_1,\tau_2}=D_{\tau_2,\tau_1}^* \;\; \tau_i=\pm,0$

\beqn
D_{+,-}=\frac{1}{2} (1-\cos^2 \theta^*) e^{2i\phi^*}& &
D_{\la,0}=-\frac{1}{\sqrt{2}} (1-\la \cos \theta^*)\sin \theta^* e^{i\la \phi^*} \\ \nonumber
D_{\la,\la}=\frac{1}{2} (1-\la \cos \theta^*)^2& &D_{0,0}=\sin^2 \theta^*
\eeqn

\subsection{Helicity amplitudes for $\gam \ra 
W^+W^-$ due to the anomalous couplings}                 
For completeness we give the helicity amplitudes for both the 
coupling $\Delta \kappa_\gamma$  that emerges within 
the effective operators we have studied, as well as the coupling 
$\lambda_\gamma$. The latter contributes also to the quartic $\gamma \gamma 
W W$
vertex. We also keep the quadratic terms in the anomalous couplings. 

The reduced amplitude ${\cal N}^{ano}_{\lambda_1 \lambda_2 \la_- 
\la_+}$  
is defined in the same way as the    \sm 
ones, {\it i.e.} in Eq.~\ref{eq:mn}, ${\cal N}^{sm}\ra{\cal N}^{ano}$
 
 \begin{eqnarray}
 {\cal N}^{ano}_{\hppll} &=&
  \dk(\gamma\sinsq+4\cossq) +4\lag\sin^2\theta+\klam(1-3\cos^2\theta)\nonumber\\
& &\frac{\dk^2}{2 }(\gamma\sin^2\theta-1+3\cos^2\theta)+
\frac{\lag^2}
{4}(1+\cos^2\theta)( \gamma\sinsq+4\cossq-2) \nonumber\\
{\cal N}^{ano}_{\hpmll} &=& -\biggl\{
 4 \dk+\frac{\kac }{4}(\gamma+2)+
 \frac{\klam}{2}(\gamma-2)+ 
 \frac{\lac}{4} [(\gamma-4)\cos^2\theta+2] \biggr\} \sin^2\theta\nonumber
\end{eqnarray}

\begin{eqnarray} 
 {\cal N}^{ano}_{\hpplt} &=& \frac{\cos\theta\sin\theta}{\sqrt{2\gamma}}\biggl\{ 
-\dk\left[ \gamma\beta+\la_-(\gamma-4)\right]-\lag\left[
\gamma\beta+\la_-(\gamma+4)\right]-\nonumber\\
& &\frac{\dk^2}{2}\left[ \gamma\beta+\la_-(\gamma-2)\right]
+\frac{\lag^2}{8}\left[\gamma(\la_--\beta)(\gamma-4)\sin^2\theta+8\la_-  \right]
-\nonumber\\ & &
\klam \la_-(\gamma+2)\biggr\}\nonumber\\
 {\cal N}^{ano}_{\hpmlt} &=& \frac{\sin\theta}{\sqrt{2\gamma}}
(1+\la_-\cos\theta)\biggl\{
  \dk (\gamma+4)+ \lag(\gamma-4) +\nonumber\\
& &\frac{\klam}{4} (\gamma-2)
(\gamma-(\gamma-4) \cos\theta\la_-) + 
 \frac{\kac}{4}\biggl\{3\gamma-(\gamma-4) \la_-\cos\theta 
\biggr\}+\nonumber\\
& &\frac{\lac}{8}\bigl[ \gamma(\gamma-2)(1-\la_-\cos\theta)^2+
 2\la_-\gamma\cos\theta(1-\la_-\cos\theta)+8\cossq\bigr]
\biggr\} \nonumber\\
 \end{eqnarray} 
 \begin{eqnarray} 
 {\cal N}^{ano}_{\hpptt} &=& 
2\dk  \ppout\left[2+\beta\la_-(1+\cos^2\theta)\right]+
\nonumber\\
& &+ \lag \biggl\{ \sin^2\theta(\gamma-2)-\ppout
\left[ \beta\la_-\left(\gamma\sin^2\theta+
2(1+\cos^2\theta)\right) 
 -4\cos^2\theta\right]\biggr\}+\nonumber\\
 & &\frac{\dk^2}{8}\biggl\{\ppout\left[\gamma\sin^2\theta-
 \beta\la_-(
\gamma\sin^2\theta-4\cos^2\theta)+6(1+\cos^2\theta)\right]+ 
 2\pmout\sin^2\theta\biggr\}+\nonumber\\
& &+\frac{\klam}{4} \biggl\{\ppout\left[
 (\gamma \sinsq+4\cossq)(1-\beta\la_-)-6\sinsq \right]+
2\pmout(\gamma-1)\sinsq\biggr\}+\nonumber\\
& &\frac{\lag^2}{16 }
\Biggl(\ppout(1-\beta\la_-)
\left[\gamma(\gamma-2)(3-\cos^2\theta)\sin^2\theta+ 
 2\cos^2\theta(5\gamma-\gamma\cos^2\theta-4)\right]\nonumber\\
&&-2\sin^2\theta(\gamma-2)(\sin^2\theta+2\ppout)-4\cos^2\theta
(\sin^2\theta+4\ppout)\Biggr) \nonumber\\
  \end{eqnarray}

 \begin{eqnarray}
 {\cal N}_{\hpmtt} &=&
 \frac{\dk}{2}\left[\ppout\sinsq+\pmout(1+\la_-\cos\theta)^2\right]
  +\lag\ppout(\gamma-4)\sin^2\theta+\nonumber\\ & &
\frac{\kac}{8}\Bigl\{6\ppout\sinsq+\pmout(1+\la_-\cos\theta)^2
(\gamma+2-(\gamma-4)\la_-\cos\theta)\Bigr\}  
 +\nonumber\\
& &\frac{\klam}{ 16}\Bigl\{2\ppout\sinsq(\gamma-3)+\pmout(1+\la_-\cos\theta)^2
(\gamma-2-(\gamma-4)\la_-\cos\theta)\Bigr\}  
 +\nonumber\\
 & &\frac{\lac}{16}\Biggl\{ 2\ppout\sinsq[2-\sinsq(\gamma-4)]+
 \pmout(1+\la_-\cos\theta)^2 \times \nonumber\\& &\biggl[
\gamma^2(1-\la_-\cos\theta)(3-\la_-\cos\theta)
-4(1-6\la_-\cos\theta+2\cos^2\theta)- \nonumber\\&&
2\gamma
(6-11\la_-\cos\theta+3\cos^2\theta)\biggr]
   \Biggr\} 
 \end{eqnarray} 

\noi
where  $P^{\pm}_{34}= 
(1\pm\la_-\la_+)/2 $ are operators projecting 
onto the W   states with same (opposite) helicities respectively.
The amplitudes not written explicitly above are simply obtained
with the relation,
\beq
{\cal M}_{\lambda_1\lambda_2;\la_-\la_+} =
{\cal M}^*_{-\lambda_1-\lambda_2;-\la_--\la_+} \;\;\;\;
\theta\ra -\theta 
\eeq

\noi
Apart for various signs due to a different
labeling of the polarisation vectors,  these amplitudes
agree with those of Yehudai{\cite{Yehudai}} save for the
dominant term (in $s_{\gamma\gamma}$) in the $\lambda_\gamma^2$ term 
of the $J_Z=0$ amplitude for transverse W's, 
{\it i.e.} 
$(1-3\cos^2\theta)\ra(3-\cos^2\theta)$. This is probably just a misprint.

\section{Properties of the helicity amplitudes for \eewwt} 
\setcounter{secnumdepth}{2} 
\setcounter{equation}{0}

The helicity amplitudes for \eewwt in the presence of tri-linear 
couplings have been derived repeatedly by allowing for all possible tri-linear couplings. 
For our purposes we will take the high energy limit 
($s\geq M_W^2, M_Z^2$). Moreover, to easily make  transparent 
the weights of the different operators in the different 
helicity amplitudes we will further assume $sin^2\theta_W \simeq 1/4$. This 
can also help
explain the order of magnitudes of the limits we have derived on the 
anomalous couplings from our maximum likelihood fit 
based on the {\em exact} formulae. Keeping the same notations as those 
for \gag but with the electron polarisation being denoted by $\sigma/2$ with 
$\sigma=\pm$ ($-$ is for a left-handed electron) and the fact that the electron and positron have opposite  
helicities, we obtain the very compact formulae for the  helicity amplitudes 
due to the chiral Lagragian parameters. $\theta$ is the angle between the $e^-$ and 
$W^-$. 
For the anomalous, the helicity amplitudes may be approximated as
\beqn
\label{eewwanoapp}
{\cal M}^{{\rm ano}}_{-; 0 0} &\sim 4 \pi  \alpha \times &\;\gamma \;\sin\theta\; 
\left\{ l_9+ 4\tilde{l}_1 +\frac{1}{3} r_9 \right\} \nonumber \\
{\cal M}^{{\rm ano}}_{+; 0 0} &\sim 4 \pi  \alpha \times  &\; \gamma\;\sin\theta\; 
\left\{ \frac{2}{3} r_9 \right\} \nonumber \\
{\cal M}^{{\rm ano}}_{-; \la 0} &\sim 4 \pi  \alpha \times &\; \sqrt{\frac{\gamma}{2}} \;
\frac{(1-\la \cos\theta)}{2} \; 
\left\{\frac{10}{3} l_9 + 4\times 2 \tilde{l}_1 +\frac{2}{3} r_9 
+\la g_5^Z\right\} \nonumber \\
{\cal M}^{{\rm ano}}_{+; \la 0} &\sim 4 \pi  \alpha \times &\sqrt{\frac{\gamma}{2}} \;
\frac{(1+\la \cos\theta)}{2} \; 
\left\{\frac{4}{3}\left( l_9 +r_9 \right) +\la g_5^Z \right\}\nonumber \\
{\cal M}^{{\rm ano}}_{-; \la \la} &\sim -4 \pi  \alpha \times & \; \sigma \; \sin\theta\; \left\{ \frac{4}{3} l_9 \right\}
\eeqn

As for the tree-level \sm amplitudes within the same approximations one has
\beqn
{\cal M}_{-; 0 0}^{\smx} &\sim -4 \pi  \alpha \times &\;\sin\theta\; \left\{ \frac{14}{3} \right\}  \nonumber \\
{\cal M}_{+; 0 0}^{\smx} &\sim -4 \pi  \alpha \times &\;\sin\theta\; \left\{   \frac{2}{3} \right\}\nonumber \\
{\cal M}_{-; \la -\la}^{\smx} &\sim 4 \pi  \alpha \times &\; 2 \lambda \frac{\sin\theta (1-\la \cos \theta)}
{1-\cos \theta)}
\eeqn
The remaining amplitudes all vanish as $1/\sqrt{\gamma}$ or faster, for example one has
\beqn
{\cal M}_{-; \la 0}^{\smx} &\sim -4 \pi  \alpha \times &\;2\sqrt{\frac{2}{\gamma}} 
\left\{ 
\frac{5}{3} -\frac{1+\la}{1-\cos \theta}\right\}\;(1-\la \cos \theta)  
\eeqn

There are a few important remarks. 
Most importantly,  both for a left-handed electron as well as for a right-handed  electron 
there is {\em effective} interference in the production of two longitudinal $W_L$'s, since 
the {\em enhancement factor} $\gamma$ does not drop out 
in ${\cal M}_{\mp; 0 0}^{\smx} \times {\cal M}^{{\rm ano}}_{\mp; 0 0}$. This also means that 
this enhancement factor will be present even in the total cross section for \eewwt. Since 
$g_5^Z$ affects primarily $W_L W_T$ production it only comes with the enhancement factor 
$\sqrt{\gamma}$. Moreover this factor will drop out at the level of the diagonal 
density matrix elements, {\it i.e.}, at the level of \eewwt. Thus, it will not be 
constrained as well as the $L_i$'s. In order to improve the limits on $g_5^Z$ one should 
exploit the non-diagonal elements. These non diagonal elements  also improve 
the limits on the $L_i$ by taking advantage of the large \sm amplitude associated with 
the production of a right-handed $W^-$ in association with a left-handed $W^+$ 
(${\cal M}_{-; \la -\la}$). 
We note that with a right-handed electron polarisation one would not be able to 
efficiently reach $L_{9L}$, since only $L_{9R}$ contribute to $W_L W_L$. 
Even for $L_{9R}$, 
considering that the yield with a right-handed electron is too small, statistics will 
not allow to set a good limit on this parameter. 
Nevertheless one could entertain the possibility of isolating the effect  $g_5^Z$ with 
right-handed electrons by considering a forward-backward asymmetry as suggested by 
Dawson and Valencia\cite{eeg5twopapers}. However this calls for an idealistic $100\%$ right-handed polarisation 
which, moreover leads to a small penalising  statistics. Therefore, we had better revert 
to a fit of the non-diagonal elements in the left-handed electron channel (or the 
unpolarised beams) case. \\
Considering the fact that at high energy the $L_i$ contribute preponderantly to 
${\cal M}^{{\rm ano}}_{-; 00}$, one expects any fit to give the best sensitivity on the 
combination given approximately by $(l_9+ 4\tilde{l}_1 +\frac{1}{3} r_9)$. This is well confirmed by 
our exact detailed maximum likelihood fit performed on the 4-fermion 
final state.

\end{document}